\documentstyle[epsfig,preprint,aps]{revtex}
\tighten
\begin{document}  
\preprint{
\vbox{
\hbox{TAUP-2356-96}
\hbox{\today{ }}
\hbox{ }
}}

\title{\Large\bf On the Possibility to Study Color Transparency  
in the Large Momentum Transfer   Exclusive  $d(p,2p)n$ Reaction}
\author{L.~L.~Frankfurt$^{a,c}$, E.~Piasetzky$^{a}$, 
 M.~M.~Sargsian$^{a,d}$, M.~I.~Strikman$^{b,c}$   }  
\maketitle

\begin{center}
{\em
$(a)$ School of Physics and Astronomy, Tel Aviv University, Tel Aviv  69978, 
Israel, \\
$(b)$ Department of Physics, 
Pennsylvania State University, University Park,  PA 16802, USA \\
$(c)$ Institute for Nuclear Physics, St. Petersburg, Russia \\
$(d)$ Yerevan Physics Institute, Yerevan, 375036, Armenia  
}
\end{center} 

\begin{abstract}
The deuteron  disintegration at high energies and  large angles in the
$d(p,2p)n$ reaction, is calculated in kinematical conditions where the
dominant contributions are due to soft rescatterings  of the initial and
final nucleons, which accompany the hard $pp$ reaction. The  eikonal
approximation,  which accounts for relativistic kinematics as dictated by
Feynman diagrams,  reveals the important role played by the initial and
final state interactions in the angular and momentum dependences of the
differential cross section. Based on these results, we propose a new and
effective test, at moderate energies, of the physics relevant for the
color transparency phenomenon in  hadron-initiated exclusive hard
processes. 
 
\end{abstract} 

\section{Introduction}

We have demonstrated recently~\cite{fpss} that it is possible 
to regulate the relative contributions  of different reaction 
mechanisms in the exclusive high energy hard  $pd\rightarrow p p n$ 
reaction, by selecting  specific kinematical conditions.
In that paper we discussed 
this reaction within the kinematics that is preferential 
to the study of short range nucleon correlations in the deuteron, 
the role of relativistic effects in the deuteron structure and the 
effects of nucleon binding.
The aim of the present paper is to analyze the reaction in the kinematics 
favorable  for the search of the Color Transparency (CT) phenomenon.

It has been shown in refs.\cite{fpss,fmgss} that, under certain  
kinematical conditions, initial and final state  interactions can 
dominate  the  cross sections.  
Those same kinematics  can be advantageous for   studying 
CT by the
vanishing initial (ISI) and final (FSI) state 
interactions of hadronic products in the  hard exclusive $pN$ 
scattering.  
(For details on CT physics, see Ref.\cite{fms94} and references therein).
As a first step for the identification of 
CT it is necessary to develop a baseline 
model which accounts for the initial and final state interactions 
within the conventional eikonal approximation. 
Thus, in this paper,  we first restrict our considerations by 
calculating the elastic rescatterings of nucleons only.
Next we will include the inelastic production  
in  the intermediate states in terms of CT.

In Ref.\cite{fmgss} the high momentum transfer exclusive deuteron
electro-disintegration reaction  has been calculated within the eikonal
approximation. The calculation of the Feynman diagrams, relevant 
for the propagation of knocked out energetic nucleons through the nuclear 
medium, revealed  certain effects related to the finite longitudinal 
distances, which are beyond the formulae of conventional 
Glauber approximation\cite{glauber}.
One of the aims of this work is to extend the formalism of  
Ref.\cite{fmgss} to account for  similar effects in the 
$d(p,2p)n$ reaction.

Another motivation of our study is to understand
the unexpected energy dependence of the
 semi-inclusive  $A(p,p)X$ reaction.
In the energy range up to the $10~GeV$, Carroll et al. Ref.\cite{carroll} 
found  that, in  semi-inclusive $A(p,2p)X$ reactions,
the  absorption of protons was reduced compared to  the value predicted 
in the Glauber approximation. At incident proton energies larger 
than $10~GeV$  an increase in the absorption was observed. 
That result  led to a  number of theoretical suggestions, such as
 the possible importance of  the interference between interactions 
of  point like- and large size configurations \cite{RP},  the 
longitudinal momentum dependence of nuclear transparency at 
pre-asymptotic energies\cite{JM}, the possible existence of a 
threshold for charmed  particle production \cite{BT}, etc. 
Exploring the deuteron in the above energy range and 
restricting the kinematics to where the deuteron wave function is well 
known, may help to elucidate the physics of hadron propagation 
through the nuclear medium and the energy dependence of the 
semi-inclusive reaction.
 
As a definition for  hard  $pp$ collision in the exclusive $d(p,2p)n$ 
reaction we   choose the conditions of large center of mass angle 
($\ge 60^0$)  $pp$ scatterings with $|t|$, $|u|$ $\ge~2~(GeV/c)^2$.
We restrict the calculations to 
momenta of the incoming proton $p_1\ge 6~GeV/c$, where a
reliable parameterization for hard $pp$ scattering exists
(see e.g.\cite{fpss}).

We deduce eikonal formulae of the scattering process 
by analyzing  the Feynman diagrams corresponding to the 
single and double rescatterings(see Fig.1).
At the present kinematical conditions the nonrelativistic 
deuteron wave function has been used but the full relativistic
kinematics of the interaction is taken into account by
the Feynman diagram approach.
At intermediate energies we 
account for relativistic kinematics exactly, i.e. we
keep also terms $\sim {1\over \sqrt{s}}$ whose contributions are  enhanced 
due to the steep momentum dependence of the deuteron wave function.
We make the approximation that the  momentum of the fast nucleon
is almost conserved in  elastic high energy soft(small angle)
collisions. Thus,  the hard process kinematics are not strongly influenced
by ISI and FSI (see Appendix B).

In section 2 we outline the details of the calculation, and 
explain the 
difference between our results and 
the nonrelativistic  Glauber approximation.  
The results can be used to calculate  nucleon 
rescatterings off any nucleus.  For the  deuteron  we 
calculate analytically the scattering amplitudes, using the known 
(see e.g. Refs.\cite{paris,bonn,BJ})
parameterization of the  wave function in  momentum space. 

In  section 3 we present the results of numerical calculations for the 
kinematical conditions that  can be  reached in the EVA experiment 
at BNL\cite{EVA}. The results reveal a strong impact of  
initial and final state interactions on the angular and momentum 
dependences
of the differential cross sections.  A surprising prediction is the 
azimuthal dependence of the final state interaction, which is a
consequence of the interplay between hard $pp$ scattering and soft  
$pn$  rescatterings.  We analyze also the range of 
applicability of the  commonly used  factorization of  
hard and soft  
scattering amplitudes.
In section 3 we  estimate the size of expected CT effects and  
suggest new options for the search for various consequences of 
CT. The simultaneous observation of all effects will make it possible 
to reach unambiguous conclusions about the onset of CT.
The results will be summarized in  section 4.

\noindent In Appendix A the analytical calculation of the rescattering 
amplitude for the deuteron is presented. 
In Appendix B we estimate the distortion of the kinematics of hard
processes due to soft $NN$ rescatterings. 

\section{Kinematics and approximations.}

In this section we calculate the cross section for the 
$d(p,2p)n$ reaction in the kinematics where the deuteron is at rest and 
the $pp$ scattering is "hard".
The kinematics of the reaction is determined by the four-momenta 
$p_d = (E_d,p_d)$, $p_1=(E_1,p_1)$, $p_3=(E_3,p_3)$, $p_4=(E_4,p_4)$, 
and $p_s = (E_s,p_s)$ of the deuteron, incoming and  scattered protons 
and spectator neutron,  respectively (c.f. fig.1a). We define also the 
variable  ${\alpha_s} = {E_s-p_{sz}\over m_d/2}$, which is  convenient
for  the description of high energy processes. 
The quantity ${\alpha_s\over 2}$ 
can be viewed  simply as  
the fraction  of the deuteron momentum carried by  the spectator 
neutron, in the infinite momentum reference frame ( in which the deuteron 
is fast).
The indices $"t"$ and $"z"$ denote the transverse and longitudinal 
directions with respect to the incoming proton momentum $p_1$.
The light cone momentum of the target proton is $\alpha = 2-\alpha_s$. 
We will use also the spectator azimuthal angle $\phi_s$, which is the 
angle between the $(p_1,p_3)$ and $(p_1,p_s)$ planes.

We choose    
 $|t|=|(p_1-p_3)^2|\sim {s-4m^2\over2}$ and  $|t|$, 
$|u|=|(p_1-p_4)^2|$ $\ge~2~GeV^2$  in order to fulfill the requirements
of  hard  $pp$ scattering at large angles.  
 With these  kinematics, in the rest frame of the nucleus, the
transverse components of the  proton  momenta  are small 
compared to the longitudinal components i.e.:
\begin{equation}
({p_{3t}\over p_{3z}})^2\sim \theta^2_3 \approx 2{m\over p_1}\alpha \ll 1;
\ \ \ \ \ \ \ \ \ \  
 ({p_{4t}\over p_{4z}})^2\sim \theta^2_4 \approx 2{m\over p_1}(2-\alpha) \ll 1
\label{eq.2}
\end{equation}
where $\theta_3$ and $\theta_4$ are the polar angles between the scattered/ 
knocked-out protons and the  incoming proton ($\vec p_1$).

We restrict  the kinematical region to 
relatively  small spectator neutron momenta i.e.  
$\alpha_s=2-\alpha\approx 1$ and $p_{st}\leq 300-400~MeV/c$. It was shown 
in ref.(\cite{fpss}) that  vacuum ($Z$)-type 
diagrams in the  $d(p,2p)n$ reaction can be neglected under those
 conditions.
By restricting the calculation to this range of Fermi momenta we can also 
neglect the contribution from possible non-nucleonic components in the 
deuteron wave function. As a result it is safe to use  a
conventional non-relativistic deuteron wave 
function (see e.g. Refs.(\cite{paris,bonn}).
   
 Fig.1 depicts all   the relevant Feynman diagrams  for the impulse 
approximation 
and for the initial and final state interactions, 
when the soft rescatterings 
are calculated within the elastic  eikonal approximation.
The solid circles in the diagrams represent the  
hard $pp$ scattering amplitude
and the broken lines represent the amplitude of small 
angle elastic NN
scattering. 
Note that, in eikonal approximation, there can be  no 
contribution from diagrams  where a 
fast proton experiences soft rescatterings off both target 
proton and slow spectator neutron. 
This is due to  the finite 
longitudinal distance between the target proton and neutron
and the fact that the sequence of    
the projectile proton colliding  off the target 
proton and neutron,
followed by a hard $pp$ scattering 
is geometrically impossible.
Similarly, it is easy to demonstrate  that  
only diagrams of Fig.1 survive  within a eikonal approximation. 
A similar cancelation of diagrams is well known in 
nonrelativistic quantum mechanics where
infinite series over exchange by potential are reduced, in the high
energy limit, to the diagrams expressed through the 
full amplitudes of
NN scattering\cite{JE}.

\subsection{The impulse approximation (IA)}

We can classify the Feynman diagrams in Fig.1 by the number of soft 
 initial and final state interactions. The simplest 
amplitude is the impulse approximation amplitude of Fig.1a. that  
can be written as:
\begin{equation}
F_a = (2\pi)^{{3\over 2}}\psi(p_s^z,\vec p_t) A^{hard}(s,t),
\label{eq.3}
\end{equation}
where $A^{hard}$ is the amplitude of the hard $pp$ scattering
and  $\psi$ is the 
nonrelativistic deuteron wave function normalized as
$\int\psi^2(p)d^3p=1$.
We are interested  in the cross 
section on a deuteron normalized to the $pp$  
cross section at the same $s$ and $t$, so that the hard $pp$ scattering 
amplitude is cancelled out.

\subsection{The single rescattering amplitude}

In the eikonal approximation 
the first order rescattering is described by the diagrams of Fig.1b,c,d.
The single rescattering amplitude ( Fig.~1b) can be 
parameterized as follows:
\begin{equation}
F_{b}=  {1\over \sqrt{2m}}\int 
{A^{hard}(s',t) \Gamma(p_d,p_s) f^{pn}(p_4,p'_s,p_s)\over 
[p_2^2-m^2+i\epsilon] [p'^2_s-m^2+i\epsilon][(p_4+ p_s - p'_s)^2 - m^2 + 
i\epsilon] } {d^4p'_s\over i(2\pi)^4},
\label{eq.4}
\end{equation}
where, in the intermediate state, $p'_s$ is the momentum of the
spectator neutron, $s' = (p_3+p_4+p_s-p'_s)^2$, 
$\Gamma(p_d,p_s)$ is the invariant vertex of the 
transitions $d\rightarrow pn$ into two off-mass shell nucleons and
 $f^{pn}$ is the  spin dependent   
amplitude of $pn$ soft scattering.
All  spin dependences of the target nucleons is included 
in the vertex factor.  
We factorize the hard scattering  amplitude $A^{hard}$ out of 
the integral because the  transferred momenta in soft rescattering are
negligible  compared to the transferred momentum of the  
hard amplitude (The validity of this approximation is discussed 
in Appendix B). 
It is reasonable to neglect also  the dependence of $f_{pn}$ on 
$p'_{s0}$  (see c.f. \cite{Gribov}) because, at high energies, the
 $NN$ interaction  depends only weakly on the collision energy. 
Using a non-relativistic description of the Fermi 
motion in the deuteron allows us  to evaluate the loop integral by taking the 
residue over the spectator nucleon energy in  the intermediate state i.e. 
we can
replace 
$[p'^2_s-m^2+i\epsilon]^{-1}d^0p'_s$ by $-i(2\pi)/2E'_s\approx -i(2\pi)/2m$. 
This is possible because in this case 
it is the only pole in the lower part of the $p'_{s0}$ complex plane. 
The calculation of the residue in  $p'_{s0}$  
fixes the time ordering from left to 
right in diagram Fig.1b. We introduce the  
nonrelativistic deuteron wave function as: 
$\psi(p_d-p'_s)\equiv {\Gamma^{d\rightarrow pn} \over [p^2_2-m^2+i\epsilon]
\sqrt{(2\pi)^32m}}$ (with $\int|\psi(k)|^2d^3k=1$)
(see e.g.\cite{Gribov,BC,Bert}). 
In the laboratory system the  amplitude  can be rewritten as:
\begin{eqnarray}
 F_{b} & = &  -{(2\pi)^{{3\over 2}}\over 2m}
A^{hard}(s,t)\int \psi(-p'_s)  {f^{pn}(p_4,p'_s,p_s)\over 
 [(p_4+ p_s - p'_s)^2 - m^2 + i\epsilon] } {d^3p'_s\over (2\pi)^3}  
\nonumber \\ & = &  
-{(2\pi)^{{3\over 2}}\over 2}A^{hard}(s,t) \int \psi(-p'_s) 
{f^{pn}(p_{st}-p'_{st})\over 2mp_{4z} [ p'_{sz}- p_{sz} + 
\Delta_4  + i\epsilon] } {d^3p'_s\over (2\pi)^3},
\label{eq.5}
\end{eqnarray}
where 
\begin{equation} 
\Delta_4 = (E_s-m){E_4\over p_{4z}} - (p_{st}-p'_{st}){p_{4t}\over p_{4z}}.
\label{eq.6}
\end{equation}

In the last part of eq.(\ref{eq.5}) we used energy-momentum conservation 
to express  the propagator of a proton of momentum
$p_4$  as:
\begin{eqnarray}
& & (p_4+ p_s - p'_s)^2 - m^2 + i\epsilon =   \nonumber \\
& &  2p_{4z}\left[p'_{sz}- p_{sz}  + (E_s-m){E_4\over p_{4z}} - 
(p_{st}-p'_{st}){p_{4t}\over p_{4z} } + {(p_s - p'_s)^2\over 2p_{4z}}\right]  
 \nonumber \\
& & \approx  2p_{4z}[p'_{sz}- p_{sz} + \Delta_4].
\label{eq.7}
\end{eqnarray}
We use the fact that the  energy  transferred  
in  the $pn$ rescattering is negligible compared to the total 
energy of the scattered particles so that we may neglect the term: 
${(p_s - p'_s)^2\over 2p_{4z}}$ with respect to  the other two contributions 
to $\Delta_4$.

We keep the term $(E_s-m){E_4\over p_{4z}}$ 
because it does not vanish 
with the increase of the projectile energy at fixed  
spectator nucleon momentum. We also keep the term   
$(p_{st}-p'_{st}){p_{4t}\over p_{4z}}$ which 
decreases only as ${1\over \sqrt{p_1}}$  according to eq.(\ref{eq.2}).
If the projectile energy is not extremely large, the contribution 
of this term in some kinematical conditions  can be comparable in size to
the first term of eq.(\ref{eq.6}).

The fact that the soft $NN$ scattering amplitude depends only weakly  on 
the initial energy helps to simplify  eq.(\ref{eq.5}). It is convenient to  
redefine the soft scattering amplitude: 
${f^{pn}(p_4,p'_s,p_s)\over 2p_{4z}m} \approx  f^{pn}(p_{s}-p'_{s})$,
where  $f^{pn}$ is now the scattering  amplitude normalized by 
the optical theorem $f^{pn}(k=0)=i\sigma^{pn}_{tot}$. We may neglect 
the  longitudinal momentum transfer in  $f^{pn}(p'_s-p_s)$ and  assume 
that the transferred momenta are almost transverse: 
i.e. $p_s - p'_{s}\approx p_{st}-p'_{st}\equiv k_t$.  Approximating
the transferred longitudinal momenta, as  $k_z\equiv p_{sz}-p'_{sz}=\Delta_4$  
and using the  fact that in the rescattering integral the average transferred 
momenta in the  $pn$ scattering are $\sim p_t$, we see  that the condition 
$\Delta_4\ll |p_t|$ (or $|p_s|\ll 2m$) allows us to neglect the 
longitudinal momentum transfer  in the soft scattering amplitude.
Therefore, under our  kinematical conditions, eq.(\ref{eq.5}) can be
rewritten as:  \begin{equation} 
 F_{b}   =    -A^{hard}(s,t){(2\pi)^{{3\over 2}}\over 2}\int \psi(-p'_s)  
{ f^{pn}(p_{st}-p'_{st})\over 
[ p'_{sz}- p_{sz} + \Delta_4  + i\epsilon] } {d^3p'_s\over (2\pi)^3},
\label{eq.8}
\end{equation}

The integration in eq.(\ref{eq.8}) can be  performed in coordinate space. 
Writing the Fourier transform of the deuteron wave function as:
\begin{equation}
\psi(p) = {1\over (2\pi)^{{3\over 2}}}\int d^3r \phi(r) e^{-ipr} 
\label{eq.9}
\end{equation}
and using the coordinate space representation of the nucleon propagator:
\begin{equation}
{1\over [ p'_{sz}- p_{sz} + \Delta_4  + i\epsilon]}  = 
-i\int d z^0 \Theta(z^0) e^{ i (p'_{sz}- p_{sz} + \Delta_4 )z^0}
\label{eq.10}
\end{equation}
we obtain  the formula for the rescattering amplitude:
 \begin{eqnarray}
F_{b}   &  = &  -A^{hard}(s,t){1\over 2 i} \int{d^2 k_t\over (2\pi)^2}d^3r   
 \phi(r)   f^{pn}(k_t)\theta(-z) e^{(p_{sz}-\Delta_4)z} e^{(p_{st}-k_t)b} 
\\ \nonumber 
& = & - A^{hard}(s,t)\int d^3r \phi(r) \theta(-z)\Gamma^{pn}
(\Delta_4,-z,-b) e^{ip_{s}r}  
\label{eq.11}
\end{eqnarray}
where   $\vec r = \vec r_p - \vec r_n$, $k_t = p_{st} - p'_{st}$.
We define a generalized profile function $\Gamma$:
\begin{equation}
\Gamma^{pn}(\Delta_4,z,b) = {1\over 2 i}
e^{-i\Delta_4z}\int f^{pn}(k_t)e^{-ik_tb}  {d^2 k_t\over (2\pi)^2}.
\label{eq.12}
\end{equation}
Eq.(\ref{eq.11}) reduces to  a Glauber-type 
approximation in the limit of zero longitudinal momentum 
transfer $\Delta_4$.
The dependence of the profile function 
on the longitudinal momentum transfer  originates 
from two sources.  
According to eqs.(\ref{eq.6},\ref{eq.7}) 
the term  $(E_s-m){E_4\over p_{4z}}$ 
accounts  for the relativistic 
kinematics which follows from  the evaluation of the 
propagator of a proton with momentum $p_4$.
A similar factor was found also for 
the  $d(e,e'p)n$ reaction in ref.\cite{fmgss}.  
The second term  
$(p_{st}-p'_{st}){p_{4t}\over p_{4z}} = {k_t  p_{4t}\over p_{4z}}$ 
is due to a slight 
misalignment of the projectile momentum 
with the  directions of the outgoing protons. Even though, according to 
eq.(\ref{eq.2}), this factor is small it has a large effect on the cross 
section due to the steep momentum dependence 
of the deuteron wave function (see discussion in section 3).  
The same modified profile function, eq.(\ref{eq.12}), 
is valid for the  single rescattering amplitudes off any nucleus 
$A$\cite{FSS}. 
The unique  feature for the  deuteron is the possibility to 
calculate the rescattering amplitudes  using  an
analytical parameterization of the deuteron wave function 
(c.f. \cite{paris,bonn}).

Substituting  the deuteron wave function in  eq.(\ref{eq.8}) by the 
analytic form of eqs.(\ref{eq.b13}), (\ref{eq.b14}) and (\ref{eq.b15}),
one can perform the integration over  $p'_{sz}$ by deforming the contour of 
integration in the upper complex half-plane of $p'_{sz}$. We can then
evaluate the integral by taking residues over corresponding poles in the 
deuteron wave function. The answer (see appendix A for details) for 
the single rescattering amplitude $F_b$ (eq.(\ref{eq.8})) is:
\begin{equation}
F_{b} =  -{(2\pi)^{{3\over 2}}\over 4i }A^{hard}(s,t)\int {d^2k_t\over 
(2\pi)^2}f^{pn}(k_t) \left(\psi^\mu(\tilde p_s) - 
i \psi'^\mu(\tilde p_s)\right),
\label{eq.18}
\end{equation}
where $\tilde p_s\equiv \vec { \tilde p_s}(p_{sz}-\Delta_4, 
\vec p_{st}-\vec k_t)$,  $\psi^\mu$ is deuteron wave function defined 
in eq.(\ref{eq.b13}) and $\psi'^\mu$ is defined in
eqs.(\ref{eq.b19}) and (\ref{eq.b20}).  

We can now   calculate also the amplitude 
corresponding to the diagram of Fig.1c,  replacing 
$p_4\rightarrow p_3$. Thus 
\begin{eqnarray}
F_{c} =  -A^{hard}(s,t)\int d^3r \phi(r) \theta(-z)\Gamma^{pn}
(\Delta_3,-z,-b) e^{ip_{s}r}  
\nonumber \\
= -{(2\pi)^{{3\over 2}}\over 4 i}A^{hard}(s,t)\int {d^2k_t\over (2\pi)^2}   
f^{pn}(k_t) \left(\psi^\mu(\tilde p_s) - i \psi'^\mu(\tilde p_s)\right),
\label{eq.21}
\end{eqnarray}
where $\tilde p_s\equiv \vec { \tilde p_s}(p_{sz}-\Delta_3, 
\vec p_{st}-\vec k_t)$  and $\Delta_3$ is  described by 
eq.(\ref{eq.6}) with the substitution $p_4\rightarrow p_3$.
 
The  initial state interaction represented by diagram 
Fig1.d. is described by an
equation similar to  eq.(\ref{eq.8}):
\begin{equation}
F_{d} = -{(2\pi)^{{3\over 2}}\over 2}A^{hard}(s,t)\int \psi(-p'_s)
{f^{pn}(p_{st}-p'_{st})\over [p_{sz}- p'_{sz} - \Delta_1 + i\epsilon]} 
{d^3p'_s\over (2\pi)^3},
\label{eq.22}
\end{equation}
where $\Delta_1$ is defined  by eq.(\ref{eq.6}) with the substitution 
$p_4\rightarrow p_1$. 
In contrast to 
eq.(\ref{eq.8}) the singularity of the propagator 
in eq.(\ref{eq.22}) is now located in the 
upper part of the complex $p'_{sz}$ plane. It is easy to show that, in 
coordinate space,  this difference corresponds to the reverse sign of the 
argument of $\theta$ function in eq.(\ref{eq.10}): 
$\theta(-z)\rightarrow \theta(z)$. In  momentum space, the contour of the
$p'_{sz}$ integration can be deformed in the  lower complex semi-plane, where 
the only pole is due to singularities in the  deuteron wave function  at 
$p'_{sz}=-i\sqrt{p'^2_{st}+m^2_j}$ (see Appendix A) and the result of the
integration differs from eq.(\ref{eq.18})  by the sign of $p'_{sz}$ only. 
Thus the  amplitude for diagram 
Fig.1d is given by:
\begin{eqnarray}
F_{d} & = &  - A^{hard}(s,t)\int d^3r \phi(r) \theta(z)\Gamma^{pn}
(\Delta_1,-z,-b) e^{ip_{s}r}  
\nonumber \\
& = & -{(2\pi)^{{3\over 2}}\over 4 i} A^{hard}(s,t)
\int {d^2k_t\over (2\pi)^2}   
f^{pn}(k_t) \left(\psi^\mu(\tilde p_s) + i \psi'^\mu(\tilde p_s)\right),
\label{eq.23}
\end{eqnarray}
where $\tilde p_s\equiv \vec { \tilde p_s}(p_{sz}-\Delta_1, p_{st}-k_t)$.

\subsection{The double rescattering amplitude}
  
The second  order  rescatterings 
where both protons scatter off the spectator neutron are represented
in the diagrams of  Fig. 1e,f,g and h. 
Using  the same approach as for the  single rescattering,  the 
double rescattering  amplitude of Fig.1e 
can be written as:
\begin{equation}
 F_{e}   =    {(2\pi)^{{3\over 2}}A^{hard}(s,t)\over 8}\int \psi(-p^{(1)}_s) 
{   f^{pn}(p^{(2)}_{st}-p^{(1)}_{st})\over 
[ p^{(2)}_{sz}- p^{(1)}_{sz} - \Delta_1  + i\epsilon] } 
 {f^{pn}(p_{st}-p^{(2)}_{st})\over 
[ p^{(2)}_{sz}- p_{sz} + \Delta_3  + i\epsilon] } {d^3p^{(1)}_s\over (2\pi)^3}
{d^3 p^{(2)}_s\over (2\pi)^3}
\label{eq.24}
\end{equation}
where  $p^{(1)}_s$ and  $p^{(2)}_s$ are the spectator (neutron) momenta before
the first and second rescatterings, respectively.

Since  the deuteron wave function in  eq.(\ref{eq.24}) does not dependent 
on
$p^{(2)}_s$, the whole dependence  on
$p^{(2)}_{sz}$ is contained in the two propagators only. But the poles of 
these propagators in the $p^{(2)}_{sz}$ complex plane are located in the lower
semi-plane. Due to fast convergence of the integral, the  contour of 
integration can be moved to the upper complex semi-plane which makes the
integral equal to $0$. 
In the coordinate space this result has a simple geometrical interpretation. 
The diagram describes an interaction of incoming and scattered 
protons with a spectator neutron which would be 
located simultaneously before and after the 
target proton which is not possible geometrically. 
Indeed using the coordinate representation of the deuteron wave function 
(eq.(\ref{eq.9})) and the nucleon propagators (eq.(\ref{eq.10})), 
we can obtain the $\Theta(z)\Theta(-z)$ term in the integrand in the 
coordinate space.
 The same reasoning leads also to  zero  for the amplitudes
of  diagrams Fig 1.f, which  can be obtained from 
diagram Fig 1.e by substituting  $p_{3}\rightarrow p_{4}$.
The amplitude in diagram  Fig. 1g, 
within the approximations discussed above  takes on the form:
\begin{equation}
 F_{g}   =    {(2\pi)^{{3\over 2}}\over 8}\int \psi(-p^{(1)}_s)   
{f^{pn}(p^{(2)}_{st}-p^{(1)}_{st})
\over 
[ p^{(1)}_{sz}- p^{(2)}_{sz} +\Delta_3  + i\epsilon] } 
 {f^{pn}(p_{st}-p^{(2)}_{st})\over 
[ p^{(2)}_{sz}- p_{sz} + \Delta_4  + i\epsilon] } {d^3p^{(1)}_s\over (2\pi)^3} 
{d^3 p^{(2)}_s\over (2\pi)^3}
\label{eq.25}
\end{equation}
where 
$\Delta_3= (E^{(2)}_s-m){E_3\over p_{3z}} - (p^{(2)}_{st}-p^{(1)}_{st})
{p_{3t}\over p_{3z}}$ 
and $\Delta_4= (E_s - E^{(2)}){E_4\over p_{4z}} - (p_{st}-p^{(2)}_{st})
{p_{4t}\over p_{4z}}$ are  different from those for  single rescattering 
(eq.(\ref{eq.6})). 
However, in our kinematics  
$E_s-E^{(2)}_s\sim {|k_t|p_S\over E_s}\ll m$ 
and  $\Delta_3$, $\Delta_4$ can be taken to be the same as 
for single rescattering.
Using eqs.(\ref{eq.9}) and eq.(\ref{eq.10}) one can write 
the amplitude, in coordinate representation, as follows:
\begin{equation}
F_{(g)}  =  {1\over 2} \int d^3r \phi(r) 
\theta(-z)\Gamma^{pn}(\Delta_3,-z,-b)
\Gamma^{pn}(\Delta_4,-z,-b) e^{ip_{s}r}, 
\label{eq.26}
\end{equation}
where $\Gamma^{pn}$ is defined by eq.(\ref{eq.12}). 
The  difference between eq.(\ref{eq.26}) and the 
conventional nonrelativistic Glauber approximation is the 
presence of the terms $\Delta_3$, $\Delta_4$.
The integration in  eq.(\ref{eq.25}) over the longitudinal components of  
nucleon momenta $p^{(1)}_{sz}$ and $p^{(2)}_{sz}$ in the
intermediate states can be evaluated by the
 technique  used in Sec.2.2 
for the calculation of the single rescattering amplitude(see Appendix A).
First, we  integrate over $p^{(2)}_{sz}$ by closing the contour of
integration at one of the poles in the complex plane (they are 
located now in  different semi-planes) to obtain:
\begin{equation}
 F_{g}   =    {-i(2\pi)^{{3\over 2}}\over 8}\int \psi(-p^{(1)}_s)  
{f^{pn}(k^{(1)}_{t})\over [p^{(1)}_{sz}- p_{sz} +\Delta_3+\Delta_4 + 
i\epsilon] } 
f^{pn}( k^{(2)}_t)  {d^2k^{(1)}_s\over (2\pi)^2}{dp^{(1)}_{sz}\over 2\pi}
{d^2 k^{(2)}_s\over (2\pi)^2}
\label{eq.27}
\end{equation}
The $p^{(1)}_{sz}$ integration in eq.(\ref{eq.27}) can be performed 
in a way similar to the one used in eq.(\ref{eq.8}) for $\Delta=\Delta_3+\Delta_4$. 
Therefore, we may use the result of the calculation of single rescattering 
to obtain:
 \begin{equation}
F_{g}  =   -{(2\pi)^{{3\over 2}}\over 16} \int {d^2k^{(1)}_t\over (2\pi)^2}
{d^2k^{(2)}_t\over (2\pi)^2}  f^{pn}(k^{(1)}_t) f^{pn}(k^{(2)}_t)
\left(\psi^\mu(\tilde p_s) - i \psi'^\mu(\tilde p_s)\right),
\label{eq.28}
\end{equation}
where 
$\tilde p_s\equiv \vec { \tilde p_s}(p_{sz}-\Delta_4-\Delta_3, 
p_{st}-k^{(1)}_t-k^{(2)}_t)$.
The expression is symmetric under interchange of 
$\Delta_4\leftrightarrow\Delta_3$. Thus,  for the amplitude 
of diagram Fig1.g we get:
\begin{equation}
F_{h}= F_{g}
\label{eq.29}
\end{equation}
\subsection{The cross section in DWIA}

In the approximation where  the hard $pp$ scattering amplitude is 
factorized (Appendix B) the cross 
section of the $d(p,2p)n$ reaction can be written in a form which is
similar to the distorted wave impulse approximation (DWIA). 
Hence the cross section for $d(p,2p)n$ reaction can be  expressed 
as follows (see e.g. Ref.\cite{fpss}): 
 \begin{equation}
\frac{d^{6}\sigma^{DWIA}}{(d^{3}p_{3}/E_{3})(d^{3}p_{4}
/E_{4}) } = {1\over 2\pi}{s^{2}-4m^{2}s\over
2m\cdot |\vec p_1 |} \frac{d\sigma}{dt}^{pp}\cdot
n(p_s)  \delta(E_{s} - (M_{D}-E))  
\label{eq.30}
\end{equation}
where $ E = E_3 + E_4 - E_1 $. 
The function $n(p_s)$ describes the distorted momentum dependence of 
a nucleon in the deuteron:
\begin{equation}
n(p_s) = {|F_a+F_b+F_c+F_d+F_g+F_h|^2\over |A^{hard}|^2(2\pi)^3}.
\label{eq.31}
\end{equation}
The factor $(2\pi)^3$ follows  from the phase space factor of 
the spectator nucleon wave function.
In eq(\ref{eq.30}) we neglected the virtuality of the interacting nucleon
\footnote{This approximation is legitimate as long as we
neglect the production of intermediate inelastic states.}
and used for  
$\frac{d\sigma}{dt}^{pp}$ the experimentally measured cross section and for 
which a phenomenological parameterization is given 
in Refs.\cite{BRC,RP,fpss}.  

\section{Numerical estimates}

We wish to calculate the quantity $T$- "Transparency"
which accounts for the  effects  due to soft 
rescatterings. Traditionally (see e.g.\cite{fms93}) $T$ is defined  as
the ratio of the experimentally measured cross section to the 
calculated IA  cross section, which according to our definitions is:
\begin{equation}
T \equiv {\sigma^{DWIA}\over \sigma^{IA} } = 
{|F_a+F_b+F_c+F_d+F_g+F_h|^2\over |F_a|^2}.
\label{eq.32}
\end{equation}
For our  numerical estimates we will use the kinematics of  
on-shell $90^0$
 center of mass
angle $pp$ scattering.
This is achieved by imposing the condition 
\begin{equation}
t   = {4m^2-s\over 2}
\label{90}
\end{equation}
on $s=(p_3+p_4)^2$ and $t=(p_1-p_3)^2$:

\subsection{Calculation within the elastic eikonal approximation}

We shall call "elastic eikonal", the approximation where 
nucleons only propagate through the deuteron and soft $pn$ rescatterings 
are described by the  diffractive $f^{pn}$ scattering amplitude. 
Since in the relevant kinematics the  amplitude $f^{pn}$ is predominantly 
imaginary,  
the IA (eq.(\ref{eq.3})) and double rescattering amplitudes 
(eq.(\ref{eq.28})) are positive, while the single rescattering amplitudes 
(eqs(\ref{eq.18}), (\ref{eq.21}), (\ref{eq.23})) are negative. 
Therefore    the  terms  in eq.(\ref{eq.32}) have
different signs and,  depending on the kinematics, the soft 
rescattering terms can either  increase or decrease  the transparency $T$,  
leading to large ISI/FSI effects on  the overall cross section.
In Refs.\cite{fpss},\cite{fmgss} we demonstrated that  the ISI/FSI is largest
for perpendicular kinematics, where the polar angle of the
neutron-$\theta_s$ is almost perpendicular to the reaction axis.
In Fig.2 we show  the spectator angular ($\theta_s$) dependence of the
transparency $T$, for different spectator momenta $p_s$. The calculation 
was done by adopting the standard parameterization for the  
diffractive $pn$ scattering amplitude:
\begin{equation} 
f^{pn} =\sigma^{pn}_{tot}(i+a_n)e^{-b_n k_t^2/2}, 
\label{eq.33} 
\end{equation} 
In our kinematics, the parameter values
$\sigma^{pn}_{tot}\approx 40~mb$, $a_n\approx -0.2$ and 
$b_n\approx~8~GeV^{-2}$ were obtained by fitting the experimental 
 $pn$ scattering data\cite{WD}.  The rather weak dependence 
of $\sigma_{tot}$ and $b_n$ on the 
nucleon momentum, for energies above 
$3~GeV$, has been included.
 
The results, presented in Fig.2, confirm the 
large ISI/FSI effects in perpendicular kinematics. 
They  reveal also the importance of the above discussed modifications 
in the conventional Glauber approximation\cite{glauber}.
We find that the dominant effect is due to  non-zero transverse momenta of
the emerging fast protons ($\sim k_t{p_{3,4t}\over p_{3,4z}}$). This effect 
reduces strongly the contribution from the interference of two 
rescattering amplitudes.\footnote{The reason of this reduction is rather 
complicated.  
In the kinematics where $\alpha_s=1$ but $p_t$ is not 
very  small, large effects of rescattering are expected. 
The deuteron wave function in the rescattering amplitude depends 
on $p_t-k_t\sim 0$ so the average transferred 
momenta in the rescattering integral are $<k_t^2>\sim p^2_t$. 
The modification due to  the $\Delta_4$ and $\Delta_3$ factors 
leads to the condition that a maximal rescattering occurs at 
$\alpha_s = 1 + p_t{p_4t\over p_4z}$ and 
$\alpha_s = 1 + p_t{p_3t\over p_3z}$ for the amplitudes $F_b$ and
$F_c$ respectively. 
Since for $90^0$ cm $pp$ scattering  $p_{3t}\approx - p_{4t}$ 
these conditions separate the kinematical regions where  
rescattering contributions are maximal and therefore the maxima of $F_b$ 
and $F_c$ are at different $\alpha_s$. As a result the interference 
terms between contributions of different rescattering amplitudes 
are reduced.}
In Fig.3 we show the $p_t$ dependence of $T$ at fixed value of $\alpha_s=1$. 
It follows from this figure that in the range $p_t<200 MeV/c$  the ISI and 
FSI  significantly screen the cross section. It is small  compared to the 
cross section calculated within the IA approximation.  
The interference between  single rescattering and 
the IA term dominates  the overall scattering amplitude. 
With increasing $p_t$ $>250-300 MeV/c$   the squared terms
and the interferences between single rescattering amplitudes  
dominate and as a result $T$ may exceed $1$. 
At even higher $p_t$ 
the second order (negative) terms in eq.(\ref{eq.32})  
(e.g. interference between single and double rescattering terms) tend to 
suppress the rise of $T$, thus changing the slope of the $p_t$ 
dependence.
 
Since at  fixed $\theta_{c.m.}$, $\theta_3$ and $\theta_4$
depends on the incoming proton energy (see eq.(\ref{eq.2})), 
the formulae of  eikonal approximation 
could lead to some energy dependence of T for fixed spectator 
momentum, even if $\sigma^{pn}_{tot}$, $b_n$ were energy independent.
This dependence is more noticeable for the larger momenta of the spectator, 
since the contribution  from ISI/FSI diagrams increases with 
increasing $p_t$ (see Fig.4). In fact, we see in  Fig.4 that
$T$ does not depend much on incident beam energy.

Another nontrivial consequence of the interplay of hard $pp$ scattering 
and soft $pn$ rescattering is the dependence of FSI on the azimuthal 
angle of the spectator neutron. The effect is due to two important 
features of the reaction: 
in hard $pp$ scatterings protons are produced at small 
but finite $\theta_3$ and $\theta_4$ angles (lab) and  high 
energy soft $pn$ rescattering is characterized by an average transferred 
momentum $<k^2>\sim p^2_t$, which is perpendicular to  
the trajectory of the  protons. 
To visualize  the origin of the  azimuthal 
dependence  let us compare  the kinematics when the  spectator 
has $\alpha=1$, $p_s\approx p_t$ and 
$\phi_s=180^0$ (in plane kinematics) and when $\alpha=1$, 
$p_s\approx p_t$ and $\phi_s=90^0$ (out of plane kinematics). 
Estimating the contribution of FSI amplitude from rescattering 
diagrams  Fig.1b,c 
it is easy to show that the difference between the $\tilde p_s$ for 
eq.(\ref{eq.21}) is:
\begin{equation}
[\tilde p^2_s]^{(in \ plane)} - [\tilde p^2_s]^{(out \ plane)}\approx 
\Delta_4^2 (\Delta_3^2)\approx <k^2>{2m\over p_1} 
\sim <p_t^2>{2m\over p_1}.
\label{azim}
\end{equation}     
Since the deuteron wave function depends strongly on momentum, the  
difference in the argument of  the wave function leads to a significant 
effect in the dependence on azimuthal angle. Therefore 
the FSI interaction for out-of-plane kinematics 
is larger than for in-plane kinematics. This is illustrated in Fig.5 where  
the $\phi_s$ dependence of transparency $T$ is calculated at $\alpha=1$ for 
different values of the spectator transverse momenta $p_t$.

Let us discuss briefly the reliability of the derived picture within the 
eikonal approximation.
The effects of the order of ($\sim
{1\over \sqrt{s}}$) 
are included in the argument of the deuteron wave 
function.
In perpendicular kinematics the deuteron wave function 
is sensitive to the momentum $\tilde p\sim 0$. Thus  $\sim
{1\over \sqrt{s}}$  effects are amplified 
by the steep momentum dependence  of the deuteron wave function at 
small momenta.
In Ref.\cite{fpss} we demonstrated 
that the off-shell effects are  minimal for the 
perpendicular kinematics since the $s$ dependence of the hard $pp$ 
scattering in this case is minimal.
For the kinematics 
where soft rescatterings dominate (at  $\alpha_s\approx 1$),  the 
calculations are practically insensitive to 
relativistic effects in the deuteron wave function~\cite{fmgss}.
A practical conclusion is that the kinematical restriction 
$\alpha\approx 1$ and $p_t<350-400~MeV/c$ 
opens a possibility to investigate reliably  effects which are 
sensitive to  initial and final state interactions of 
energetic protons with the slow spectator neutron.
Another important question is the validity
of the factorization of   hard and 
soft scatterings. 
We demonstrate in Appendix B that the errors 
due to this factorization are minimal for small values of the spectator 
transverse momenta and $\alpha_s\ge 1$.
The numerical estimates of the errors 
are shown in Fig.12, which shows 
that, in the region of $0.9<\alpha_s<1.1$ and $p_t\leq 400~MeV/c$,
the factorization approximation is valid within an accuracy of better than 
$15\%$ for energies $p_1\ge 6~GeV/c$.
The accuracy improves at higher 
incident energies (as shown in  Fig.12, for calculations at  
$p_1=6~GeV/c$ and $15~GeV/c$.)

\subsection{Implication for color transparency} 
 
The aim of the paper is to explore whether
 the hard exclusive 
$d(p,2p)n$ reaction can initiate effects of color transparency. 
The main idea is that at the point of interaction the hadrons 
of the hard elastic scattering 
are in a "point like" configuration (PLC) whose subsequent
strong interaction is reduced.\cite{brod,muel,fms94,fms93}.
 As a result  soft interactions shortly before and after the hard collision 
will be lower than the usual strong interaction of hadrons.
Since a PLC, produced in the hard process,  is not an eigenstate 
of the QCD Hamiltonian  but rather a wave-packet of destructively 
interfering  mass states, it has to evolve eventually 
into a final hadron state. Due to time dilation,
the characteristic distances for the evolution  
of a PLC to the normal hadronic state increases  with the  
total energy of the PLC. 

The exclusive $d(p,2p)n$ reaction has several features which are 
very sensitive to CT. Since ISI/FSI occur at  internucleon 
distances in the deuteron of about $\sim 1~Fm$\cite{fmgss},
one can use the spectator neutron to tag the PLC at an early stage 
of evolution thus reducing expansion effects. 
As a result,   large CT effects can be expected even at 
comparatively low energies.
Large spectator momenta $p_{s}$ with  $\alpha_s\approx 1$ 
ensure small impact parameters for soft $pn$ 
rescatterings. This will enhance the  ISI/FSI contribution to 
the $d(p,2p)n$ cross section. 

We  calculate $T$ of 
eq.(\ref{eq.32}) in a model which accounts for PLC formation and 
its time development in the  framework of the quantum diffusion model 
(QDM)\cite{flfs}. 
We note that CT in  the resonance basis 
representation of PLC\cite{fgms92} predicts rather similar 
effects\cite{fmgss}.
We follow the procedure 
described in ref.\cite{efgmss94} to calculate  the $pn$ scattering amplitude 
and profile function of 
eq.(\ref{eq.12})) within QDM\cite{flfs}. The $pn$ scattering amplitude 
$f^{pn}$ in eq.(\ref{eq.11}), (\ref{eq.21}), (\ref{eq.23}), and 
(\ref{eq.26})  is replaced by a  position dependent one:
\begin{equation}
f^{PLC,N}(z,k_t,t)   =   i\sigma_{tot}(z,t) \cdot 
e^{-{b_n\over 2 }k^2_t}    {G_{N}(t\cdot\sigma_{tot}(z,t)/\sigma_{tot})
\over G_{N}(t)},  
\label{eq.34}  
\end{equation}  
where $G_{N}(t)$ is the Sachs form factor,
$\sigma_{tot}(z,t)$  is the  effective total cross section of the  
interaction  of the PLC at distance $z$ from the interaction point. 
According to ref.~\cite{flfs}:
\begin{equation}
\sigma _{tot}(z,t)    =   \sigma_{tot}^{pn}
\left \{ \left ({z \over l_{h}} + 
{\langle r_{t}(t)^{2} \rangle \over \langle r_t^{2}  \rangle } 
(1-{z \over l_{h}}) \right )\Theta (l_{h}-z) 
  +  \Theta (z-l_{h})\right\}, 
\label{eq.35}   
\end{equation} 
where ${l_h = 2p/\Delta M^{2}}$, with ${\Delta M^{2}=0.7-1.1GeV^{2}}$. 
Here ${\langle r_{t}(t)^{2} \rangle}$  represents the transverse size of the  
initially produced configuration. 
Theoretical analysis of 
realistic models of a nucleon indicate\cite{fms93} 
that this size is rather small even for  $|t|~\geq~1.5~GeV^2$. 
Any effects of interplay between large and small size 
configurations can be included in a rather straightforward
manner\cite{BT,RP}. However our aim here is primarily to 
demonstrate sensitivity to CT.

The QDM calculation should be compared with the elastic eikonal calculation 
of the previous section, where   $pn$ soft rescattering  
was  taken to be  practically  energy independent 
(see eq.(\ref{eq.33})). In order to emphasize any
color transparency effects, we calculate again the dependence of $T$ on 
the spectator 
transverse momentum, 
azimuthal angle and incoming proton momentum, since we showed 
in Sec.3.1 that all these quantities 
are rather sensitive to the soft $pn$ interactions. 
In Fig.6,  we present the $p_t$ dependence of $T$ (eq.(\ref{eq.32}))
calculated  in the eikonal 
model with 
 eq.(\ref{eq.33}) and in the QDM for the rescattering amplitude with 
eqs.(\ref{eq.34},\ref{eq.35}). The shaded areas  correspond to the range of 
the parameter $\Delta M^2 = 0.7 - 1.1~GeV^2$ which characterizes the scale of 
excitation energies in PLC and which controls the distance over which 
the PLC evolves to the normal hadronic state. Large values of $\Delta M^2$ 
correspond to small deviations from the elastic eikonal prediction.  
Fig.6 also shows that, at fixed initial energy, 
CT effects become more prominent with increasing $p_t$.  This can be 
understood by the fact that
the contribution of the IA amplitude is  small and 
the soft rescatterings occur at small internucleon distances 
leading to some suppression  in expansion of PLC. Also, 
higher 
order rescatterings are more important at larger $p_t$. They are 
proportional to higher powers of $\sigma^{tot}_{pn}(z,t)$ and 
therefore should be more suppressed by the onset 
of CT.
The dependence of $T$ on the azimuthal angle of the spectator,
in the  $\alpha=1$ kinematics, is shown in Fig.7.  The ISI 
(Fig.1d) does not contribute to the 
$\phi_s$ dependence because we chose the $z$ axis (lab)  
in the  $p_1$ direction. We see in Fig.7 that the onset of CT in the $\phi_s$ 
dependence requires larger  projectile energies. This is 
because the CT impacts the FSI only at energies $p_3\sim p_4\sim p_1/2$. 
At energies ($\sim 6~GeV$), before the onset of CT,  the
$\phi$ dependence can be used to check  predictions of the 
elastic eikonal approximation. 
 
To discuss the energy dependence of $T$ we include in our
consideration also 
the models which accounts for both the PLC and 
large (Blob) size configurations in hard scattering $pp$ amplitude.
We follow to the prescription of Ref.\cite{JM92} and represent the 
hard $pp$ scattering amplitude as:
\begin{equation} 
A^{Hard}(s,t) = A^{PLC}(s,t) + A^{BLC}(s,t).
\label{sum}
\end{equation} 
$A^{PLC}(s,t)$ is the component of hard scattering amplitude
which produces the PLC and leads to CT.  $A^{BLC}(s,t)$
corresponds to  the large size (soft) component of 
$pp$ scattering amplitude and has a  cross section 
comparable to $\sigma_{total}$.
The source of such a soft component could be either the 
opening of $c \bar c$  channels at energies near threshold\cite{BT} 
or the presence of Landshoff processes in hard $pp$
scattering\cite{RP}.
Implications of both mechanisms for CT were discussed in details in
Ref.\cite{JM92}, where PLC expansion effects are accounted for within 
the multi resonance expansion model of CT. In the present calculation 
we describe the expansion effects within QDM using the same 
Ralston-Pire parameterization for $A^{BLC}(s,t)\cite{RP}$. 
The numerical results of the multi resonance model and QDM for
CT in $(p,2p)$, and $(e,e'p)$ reactions are very close \cite{fms94}.
In Fig.8 we present the energy dependence of $T$ at two values of 
$p_{st}=0.2~GeV/c$ and  $p_{st}=0.4~GeV/c$. It shows that the higher 
the spectator transverse momentum, the larger is the  CT effect. 
The increase of the incoming energy of the proton diminishes 
the sensitivity of $T$ to $\Delta M^2$, due to  
reduced sensitivity to the expansion.    
The model which accounts for the interference between large and small
size components of hard $pp$ scattering, as expected, reveal 
oscillation with the increase of projectile energy.
New feature is that amplitude of these oscillations increase 
with $p_t$ and the phases of oscillations are opposite for 
kinematics dominated by screening ($p_t=0.2~GeV/c$) and by 
rescattering ($p_t=0.4~GeV/c$) effects of ISI/FSI.

As we have seen in Sec.3.1, $T<1$ at 
$p_{st} \lesssim  0.2 GeV/c$, and $T>1$ at $p_{st}\gtrsim 0.3~GeV/c$, 
(see, Figs.3,6,8). Since, by definition, $T=1$ means complete transparency 
(i.e. no ISI/FSI) CT  produces opposite trends  
in the regions where $T<1$ and $T>1$ (see e.g. Fig.6b). This property  
calls for a more sensitive quantity to characterize CT
\cite{fmgss,efgmss94}. We define  the ratio: 
\begin{equation}
 R={\sigma(p^{(1)}_{st})\over \sigma(p^{(2)}_{st})},
\label{eq.36}
\end{equation} 
where $p^{(1)}_{st}$ and $p^{(2)}_{st}$   are such that 
$T(p^{(1)}_{st}) >1$ and $T(p^{(2)}_{st}) < 1$. This ratio is more 
sensitive to CT due to different trends of CT effects in the numerator and  
denominator of $R$ while it will be less sensitive to uncertainties in
the theoretical calculations.
In Fig.9 we present the energy dependence of $R$, which shows 
deviations of QDM prediction from the 
eikonal approximation, in the range of the initial 
proton momenta $\ge 6~GeV/c$, by a  factor $2-3$ due
to  CT. 
Because of opposite phases of energy oscillations within the 
models which accounts for the interference between BLC and PLC
(Fig.8), $R$  reveals profound oscillations in Fig.9.
The dependence of $R$ on the azimuthal angle of the spectator - Fig.10
at fixed energies of initial proton could be used as a  complementary 
method in the  study  of CT.  Due to large  FSI in the out-of-plane 
kinematics, the QDM prediction of CT at $p_1=15~GeV/c$ could  be enhanced
by the factor of $5-6$ over the eikonal results.

It was shown in Ref.\cite{fmgss96} that IA is strongly 
suppressed for 
inelastic recoil final states in the deuteron fragmentation region. 
This was due to the fact that
under those circumstances 
the  IA is controlled  by the inelastic component 
of the deuteron
ground state, which is negligible for  recoil momenta $\le 700~MeV/c$.
Thus the cross section is dominated 
by the rescattering diagrams. Therefore 
the CT  will manifest itself by a 
reduced   production of low momentum 
inelastic states in the deuteron fragmentation region 
at increasing projectile energies. In Fig.11 we 
present the 
calculation of the ratio of 
the cross section of quasielastic $d(p,2p)n$ reaction to that of 
$d(p,2p)N^*(1580)$ reaction. The calculations are done   
in the QDM and  the elastic eikonal approximation \cite{fmgss96}.
Fig.11 reveals  a strong sensitivity of the ratio to CT.
For incident 
momenta from  $6 - 18 GeV/c$ the  eikonal approximation shows a 
drop in the ratio of $10\%$, while the inclusion of CT 
leads to  a 
decrease by a  factor of $3$. This 
suggests a very
simple method for studying CT  effects i.e.
the measurement of the energy dependence of quasielastic and inelastic 
rates in the deuteron fragmentation region. 

\section{Summary}

We calculated the cross section of hard exclusive $d(p,2p)n$ reaction 
taking into account  the initial and final state 
interactions, within the elastic eikonal approximation. 
Analysis of Feynman diagrams in the relativistic domain produced
significant effects beyond 
the  conventional nonrelativistic Glauber type formulae. 
The results can be applied to any nucleus. 
The  deuteron calculations were done analytically
using a well known 
form of the deuteron wave function.

The calculation produced a diffractive pattern in the cross section
due to ISI/FSI.
In the kinematics where $\alpha_s\approx 1$, 
$p_{st}\le 0.2~GeV/c$ the ISI/FSI suppress the cross section 
with respect to  IA,
while at $\alpha_s\approx 1$ $p_{st}\ge 0.3~GeV/c$  the ISI/FSI 
increases the cross section.  
Another result is the  prediction of an azimuthal angle 
dependence due to FSI.
This is a result of including small but 
finite angles of  the scattered protons
into the soft rescattering amplitude.
We estimated the validity of the factorization approximation within
the elastic eikonal approach.
We found that  the factorization approximation is valid for 
$\alpha_s\approx 1$ to  better than $15\%$ for $p_{st}\le 0.4~GeV/c$.

Since ISI/FSI dominate the cross section in $d(p,2p)n$, the CT
phenomenon will be best observed in this reaction.
We calculated the transparency $T$ with the quantum diffusion 
model and model where the interference between small and large size
components of large angle $pp$ scattering are taken into account explicitly. 
Exploiting the fact that CT leads to 
opposite effects in the kinematical regions
where  $T<1$ and $T>1$, we introduced 
the ratio  $R = {\sigma(T>1)\over \sigma(T<1)}$, which reveals extra 
sensitivity to CT effect. At presently accessible energies, 
 the predictions of energy and angular dependences of $R$
differ by as much as  a factor of 2-6  
between the  eikonal and CT calculations. 
The model with the interference of large and small size
configurations predicts profound oscillations for the energy 
dependence of $R$.
  
An additional possibility to search for  CT was found.
It consists of measuring the energy dependence of the ratio of the 
production rates of
elastic and inelastic recoil states in the deuteron fragmentation
region. While conventional calculation 
predict practically no 
energy dependence, the CT predicts  a significant 
drop of this ratio with  increasing the projectile energies.

\section{Acknowledgments}

We are grateful to Jonas Alster for valuable comments and 
for reading the manuscript. 
We would like to thank Jerry Miller for useful discussions and 
for providing his calculation of large angle $pp$   scattering amplitude. 
We wish to thank also Alan Caroll, Steve Heppelmann,
Yael Mardor and Israel Mardor for helpful discussions.
Two  of us (M.S. \& M.S.)  thank the DOE Institute for Nuclear Theory at
the University of Washington for its hospitality and support during 
the workshop ``Quark and Gluon Structure of Nucleons and Nuclei'', 
where part of this work was carried out.

This work was supported, in part, by the Basic Research Foundation 
administrated 
by the Israel Academy of Science and Humanities,
by the U.S.A. - Israel Binational Science Foundation  
Grant No. 9200126 and by the U.S. Department of Energy 
under Contract No.\ DE-FG02-93ER40771.

\vspace{0.4cm}

\appendix 
 
\vspace{0.4cm}

\section {Analytic calculation or rescattering amplitude}

\vspace{0.2cm}

We   calculate the rescattering amplitude 
in eq.(\ref{eq.8}) by the method described in  ref.\cite{fmgss} 
using the deuteron wave function in 
momentum space,  defined as\cite{BJ}:
\begin{equation}
\psi^\mu(p) = {1\over \sqrt{4\pi}}\left(u(p) + w(p)
\sqrt{1\over 8}S(p_z,p_t)\right)\chi^\mu
\label{eq.b13}
\end{equation}
where $\chi^\mu$ is the deuteron spin function and 
\begin{equation}
S(p_z,p_t) = {3(\vec \sigma_p \cdot \vec p)(\vec\sigma_n 
\cdot\vec p)\over p^2} -  
\vec\sigma_p\cdot\vec\sigma_n 
\label{eq.b14}
\end{equation}
where $\sigma_p$, $\sigma_p$ Pauli matrices.
The functions $u(p)$ and $w(p)$ are the radial wave functions 
of $S$- and $D$- states, respectively and they can be  written
as\cite{paris,bonn}:
\begin{equation}
u(p) = \sum\limits_{j} {c_j\over p^2 + m^2_j}; \ \ \ \ \ \ \ \ 
w(p) = \sum\limits_{j} {d_j\over p^2 + m^2_j} 
\label{eq.b15}
\end{equation}
where $\sum\limits_j c_j = \sum\limits_j d_j = 0$, which guarantees that
$u(p),w(p)\sim {1\over p^4}$ at large 
$p$  and $\sum\limits_j {d_j\over m^2_j}=0$ to provide $w(p=0)=0$. 
Insertion of  eq.(\ref{eq.b15}) 
into the eq.(\ref{eq.8}) gives: 
\begin{eqnarray}
& &  F_{b}  =   -{(2\pi)^{{3\over 2}}\over 2}\sum\limits_j 
A^{hard}(s,t)\int {d^2p'_{st}\over (2\pi)^2} f^{pn}(k_t) 
\nonumber \\
& & \times\int {d p'_{sz}\over (2\pi)}
\left({c_j\over p'^2_s + m^2_j} + {d_j\over p'^2_s + m^2_j}\sqrt{1\over 8}
S(p'_{sz},p'_{st})\right)
{\chi^\mu  \over  p'_{sz}- p_{sz} + \Delta_4  + i\epsilon } 
\label{eq.b16}
\end{eqnarray}  
Substituting    
$p'^2_s + m^2_j = (p'_{sz} + i\sqrt{m_j^2+p'^2_{st}})(p'_{sz} - 
i\sqrt{m_j^2+p'^2_{st}})$
one can perform the integration over $p'_{sz}$ by closing the 
contour in the upper 
$p'_{sz}$ complex semi-plane.
Note that the $p^{-2}$ dependence of the tensor function $S(p)$ will not 
introduce a new 
singularity, since $w(p=0)=0$. Setting the 
residue at the point $p'_{sz} = i\sqrt{m_j^2+p'^2_{st}}$ 
we obtain:
\begin{eqnarray}
& & F_{b}  =   -{i(2\pi)^{{3\over 2}}\over 2}\sum\limits_j 
A^{hard}(s,t)\int {d^2p'_{st}\over (2\pi)^2} f^{pn}(k_t)  
\left[ {c_j\over 2i\sqrt{p'^2_t + m^2_j}} + \right. \nonumber \\
& &  \left. + {d_j\over 2i\sqrt{p'^2_t + m^2_j}}\sqrt{{1\over 8}} 
S(i\sqrt{p'^2_t + m^2_j},p'_{st})\right]
{\chi^\mu  \over  i\sqrt{p'^2_t + m^2_j}- p_{sz} + \Delta_4}. 
\label{eq.b17}
\end{eqnarray}
After regrouping  of the real and imaginary parts, the  above equation can be 
rewritten as:
\begin{equation}
F_{b}  =   -{(2\pi)^{{3\over 2}}\over 4i  } A^{hard}(s,t)
\int {d^2k_t\over (2\pi)^2} f^{pn}(k_t) 
\left(\psi^\mu(\tilde p_s) - i \psi'^\mu(\tilde p_s)\right),
\label{eq.b18}
\end{equation}
where $\tilde p_s\equiv \vec { \tilde p_s}(p_{sz}-\Delta_4, 
\vec p_{st}-\vec k_t)$, $\psi^\mu$ is the wave function defined in 
eq.(\ref{eq.b13}) and $\psi'^\mu$ is defined as:
\begin{equation}
\psi'^\mu(p) = \left(u_1(p)p_z + {w_1(p)p_z\over \sqrt{8}}S(p_z,p_t) + 
{w_2(p)\over \sqrt{8} p_z}\left[S(p_z,p_t) - S(0,p_t)\right]\right)\chi^\mu,
\label{eq.b19}
\end{equation}
where 
\begin{eqnarray}
u_1(p) &  = & \sum\limits_j{c_i\over \sqrt{p^2_t+m^2_j}(p^2+m^2_j)},
\ \ 
w_1(p) = \sum\limits_j{d_i\over \sqrt{p^2_t+m^2_j}(p^2+m^2_j)}, \nonumber \\ 
w_2(p) &  = & \sum\limits_j{d_i\over \sqrt{p^2_t+m^2_j}m^2_j},
\label{eq.b20}
\end{eqnarray}
Note that the last term in eq.(\ref{eq.b19}) does not have a
singularity at $p_z=0$ since $(S(p_z,p_t) - S(0,p_t))\sim p_z$.

\vspace{0.4cm}

\section{Validity of the factorization approximation}

\vspace{0.2cm}

In this Appendix,  we check the validity of factorization of 
the  hard 
$NN$ scattering amplitude from the integrals over the soft rescattering. 
In the rescattering amplitudes (eqs.(\ref{eq.18}), (\ref{eq.21}), 
(\ref{eq.23}), (\ref{eq.28}), (\ref{eq.29})) $s'$ and $t'$ which 
enter in the hard amplitude $A^{hard}_{pp}(s',t')$
differ from the measured $s=(p_3+p_4)^2$ and $t=(p_1-p_3)^2$. 
For the  rescattering diagrams we have:
\begin{eqnarray}
\begin{array}{lll}
 s'= (p_3+p_4+k)^2 \ \ \ \ &  t' = t \ \ \ \ & 
\mbox{(Fig.1b)} \\ 
 s'= (p_3+p_4+k)^2 \ \ \ \ &  t' = (p_1-p_3-k)^2 \ \ \ \  & 
\mbox{(Fig.1c)} \\ 
 s'= s             \ \ \ \ &  t' = (p_1-p_3-k)^2 \ \ \ \  & 
\mbox{(Fig.1c)} \\ 
\end{array}
\label{ntrue}
\end{eqnarray}
where $k$ is the transferred momentum  in soft rescattering.
To estimate the difference between $s'$, $t'$ and $s$, $t$ 
it is important to take into account the specific feature of 
soft,  high energy collisions, i.e. that the transferred momenta are 
predominantly transverse 
with respect to the  momentum of fast projectile 
i.e.  $\vec p_j\vec k=0$,( $j=1,2,3$). The effects 
neglected within the factorization approximation can be estimated 
by using the  hard $pp$ scattering amplitude in the form 
$A^{pp}(s,t)\sim {1\over s^{0.8} t^{3.2}}$ (c.f. Ref.\cite{BRC}).  
With this parameterization 
we obtain:
\begin{eqnarray}
\chi = {A(s',t')\over A(s,t)}\approx
 \left\{  
\begin{array}{lll}
\left(1 - {0.8<k^2>\over 4 s}
{4(2-\alpha_s)-\sqrt{\alpha_s(2-\alpha_s)-1}\over 2-\alpha_s}\right)
& \mbox{(Fig.1b)} \\
\left(1 - {0.8<k^2>\over 4 s}
{4(2-\alpha_s)-\sqrt{\alpha_s(2-\alpha_s)-1}\over 2-\alpha_s}\right)
(1 - {12.8<k^2>\over  s})
& \mbox{(Fig.1c)} \\
\left(1 - {12.8<k^2>\over  s}\right)
& \mbox{(Fig.1d)} 
\end{array}
\right.
\label{est}
\end{eqnarray}
In deriving  eq.(\ref{est}) we use eq.(\ref{eq.2})
for the scattering angles in hard $pp$ scattering 
and average over the direction of  
transferred   momenta  at soft rescatterings (as a result 
the terms proportional to $\vec k$ vanished). 
Using for  $\alpha_s=1$,  the relation $<k^2>\sim p^2_{st}$ in high 
transverse momentum of the spectator nucleon  we  estimate $|\chi-1|$ 
$\sim 15\%$ at $p_1=6~GeV/c$ and $\sim 6\%$ at $p_1=15~GeV/c$.
The overall effect in the full amplitude is smaller, 
because the 
contribution of the impulse approximation (diagram of Fig.1a) 
does not contribute to the error.
Eq.(\ref{est}) shows that uncertainties due to the factorization 
approximation are minimal at small transverse spectator  momenta 
(due to the relation $<k^2>\sim p^2_{st}$). 
Also, for fixed spectator momenta, the 
uncertainties are smaller 
at $\alpha_s< 1$.
Despite  the fact that the ratio in eq.(\ref{est}) 
increases at $\alpha_s>1$ 
the error due to 
factorization in the overall amplitude will still be small
since, in these cases, the  IA amplitude 
dominates (see e.g. Fig.12 below and 
Ref.\cite{fpss}). The reason for the  dominance of IA 
in $\alpha_s>1$  kinematics 
is that,  according to eq.(\ref{eq.6}), 
the argument of the deuteron wave function in IA is smaller 
than the one in rescattering amplitude.
The  error due to 
factorization in the double rescattering diagrams of Fig.1g,h is even 
less important since, in our kinematics, they are just a  
correction to single rescatterings (see Fig.2).

To demonstrate  numerically the accuracy of factorization,
we use two models to describe the $A^{pp}(s,t)$ amplitude. In the first model 
we use the fit to hard exclusive $pp$ cross section $\sim s^{-10}
[1-cos^2(\theta_{cm})^{-4\gamma}]f(t,s)$,  where $f(t,s)$ is a slowly 
varying function of $s$ and $t$, (see for details Refs.\cite{RP} 
and \cite{fpss}),) for  the c.m. scattering angles $\theta_{c.m.}\ge 60^0$,
and we  assume a  $s^{2(\alpha(t)-1)}$ dependence for smaller angles. 
In the second model we used a power law dependence 
$\sim s^{-10}[1-cos^2(\theta_{cm})^{-4\gamma}]f(t/s)$  down to 
$|t|\approx 1 GeV^2$ and  for smaller $t$ we used the ordinary diffraction 
formulae, as in eq.(\ref{eq.33}). The results are given in Fig.12, 
which shows that in the region of $0.9<\alpha_s<1.1$ and 
$p_t\leq 400~MeV/c$ the factorization approximation is valid 
within $\le 15\%$ at $p_1=6~GeV/c$ and even less for higher energies.

\vspace{0.5cm}
 

\begin{figure}[p]
\centerline{
\epsfig{file=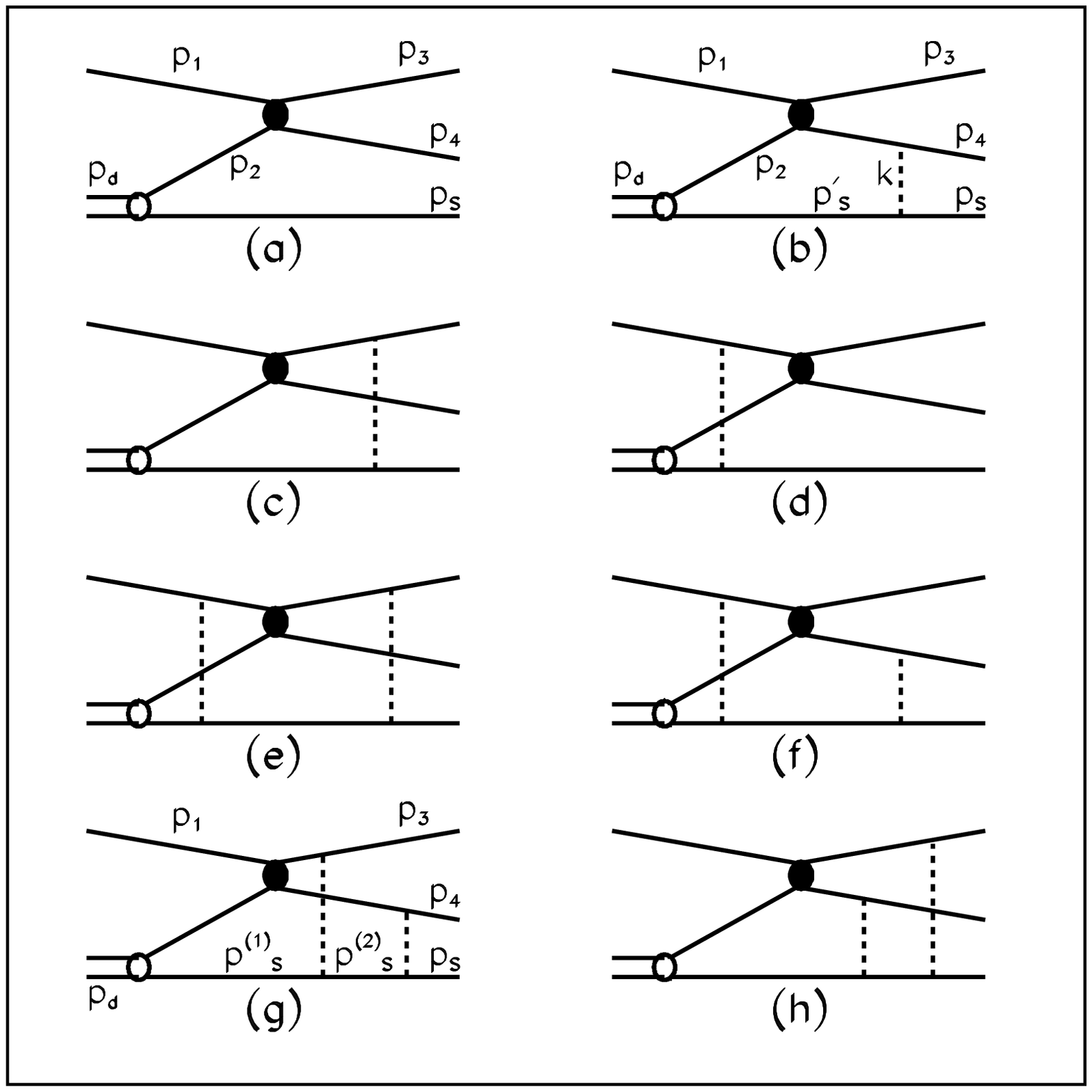,width=14cm,height=14cm}
}
\caption{ {\em Feynman diagrams of the  eikonal approximation for $d(p,pp)n$ 
scattering. The dashed lines describe the amplitude of $NN$ scattering, 
the full  circles represent the hard $pp$ scattering  amplitude.}} 
\end{figure}

\begin{figure}[p]
\centerline{
\epsfig{file=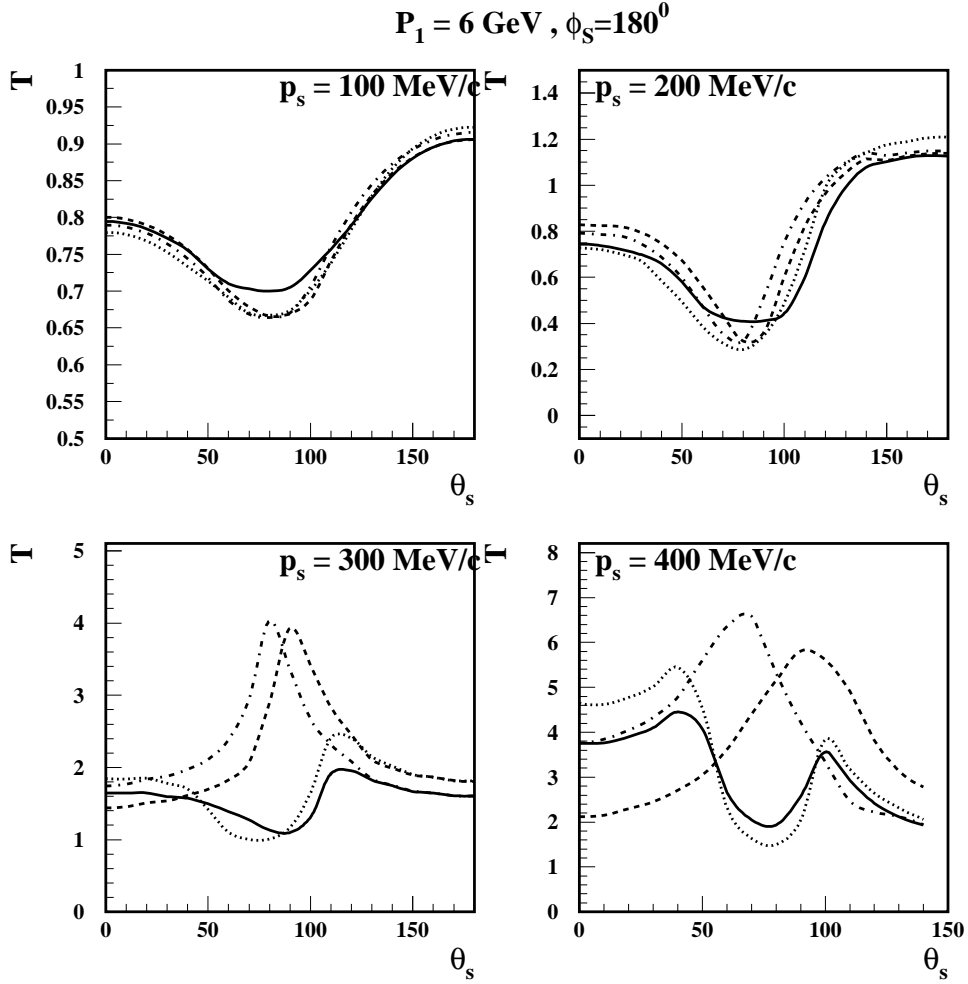,width=14cm,height=14cm}
}
\caption{ {\em 
Transparency $T$ as a function of angle $\theta_s$ 
between the momentum of spectator neutron and projectile 
proton. The dashed lines are for
the  conventional Glauber approximation where 
$\Delta_3= \Delta_4 =\Delta_1=0$ 
in the rescattering amplitude. The dash-dotted are 
the  calculations which include the  relativistic correction 
$(E_s-m){E_{3,4}\over p^z_{3,4}}$ in $\Delta_3$  and $\Delta_4$. 
The solid lines include the 
transverse momentum of the protons in  $\Delta_3$  and $\Delta_4$ 
(e.g. eq.(6)). The dotted lines are the same as  the
 solid 
lines, but without double rescattering terms.}} 
\end{figure}

\begin{figure}[p]
\centerline{
\epsfig{file=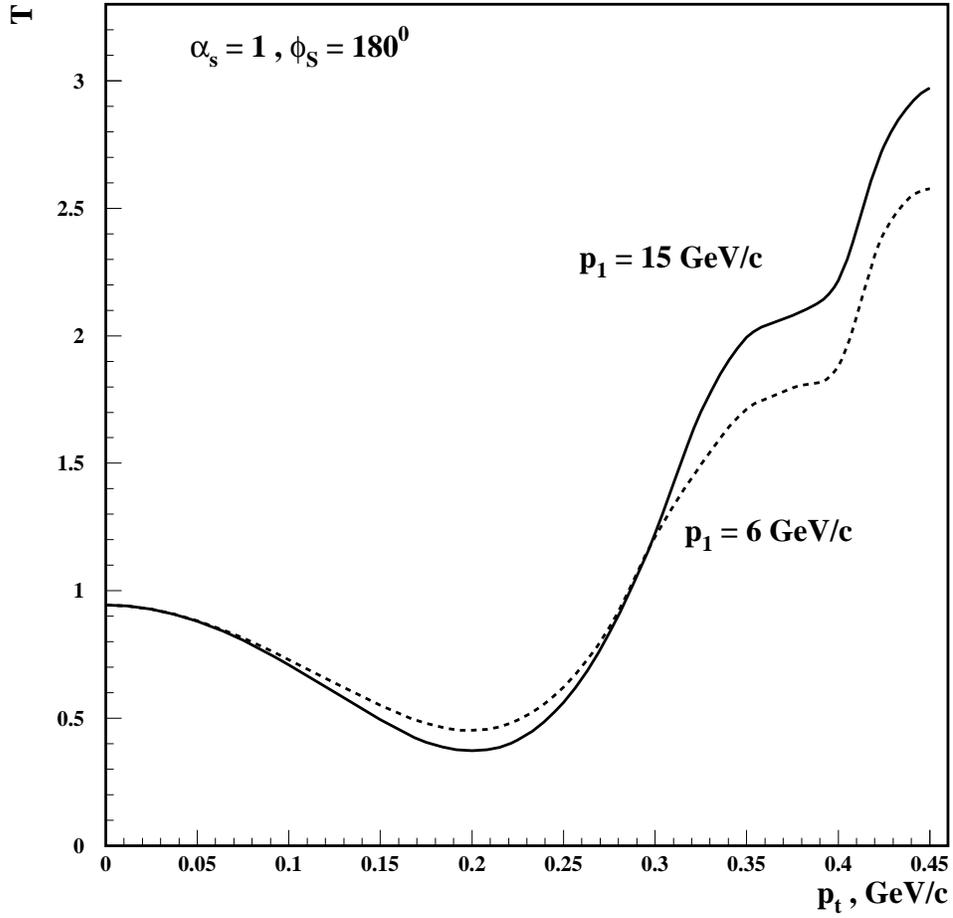,width=14cm,height=14cm}
}
\caption{{\em The $p_t$ dependence of the transparency $T$ at $\alpha_s=1.$  
The dashed line corresponds to $p_1=6~GeV/c$ and solid line 
 to  $p_1=15~GeV/c$.}}
\end{figure}

\begin{figure}[p]
\centerline{
\epsfig{file=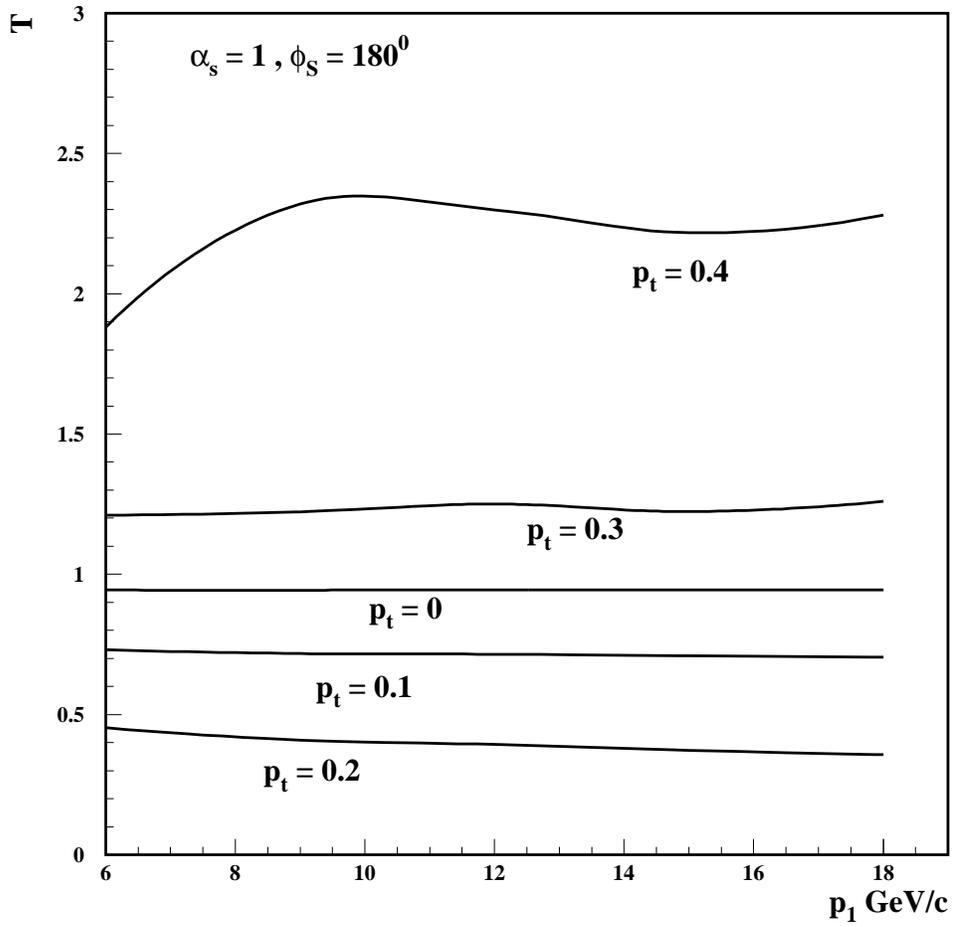,width=14cm,height=14cm}
}
\caption{ {\em  The $p_1$ dependence of the transparency $T$ at $\alpha_s=1$
for various $p_t$.}}
\end{figure}

\begin{figure}[p]
\centerline{
\epsfig{file=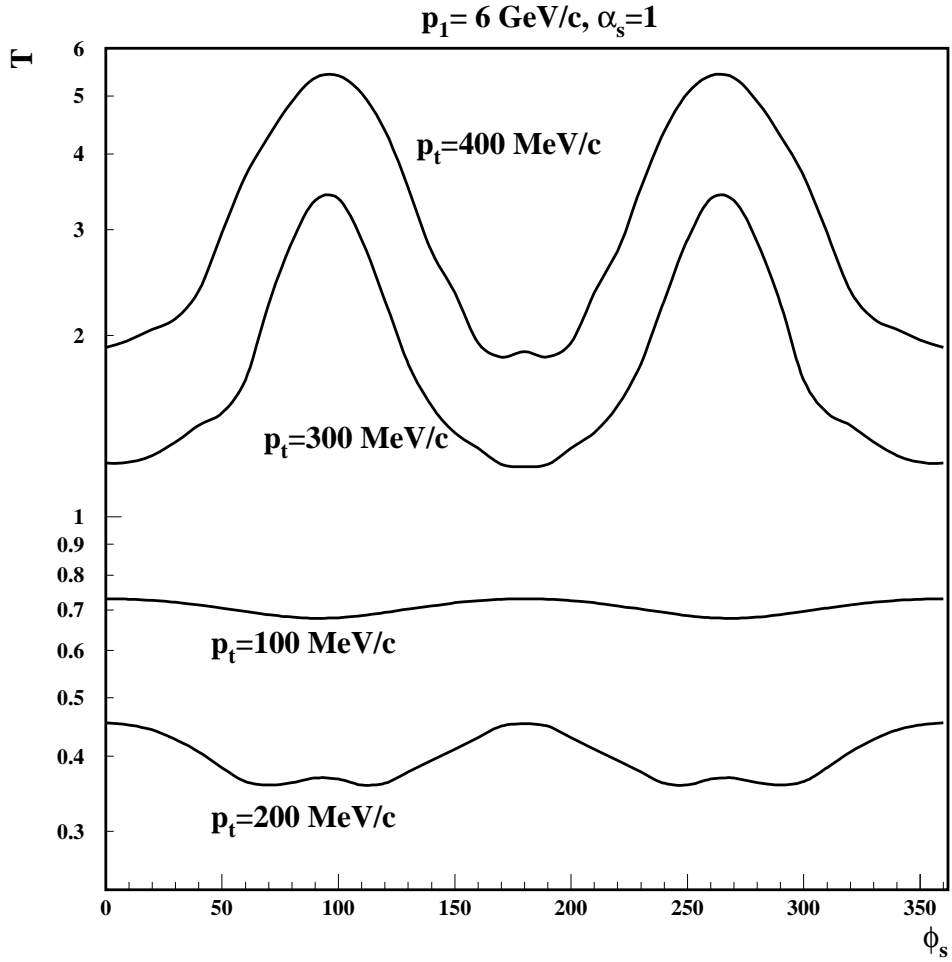,width=14cm,height=14cm}
}
\caption{ {\em The  $\phi_s$ dependence of $T$,
for $\alpha_s=1$ and $p_1=~6~GeV/c$, for different values of $p_t$.}}  

\end{figure}

\begin{figure}[p]
\centerline{
\epsfig{file=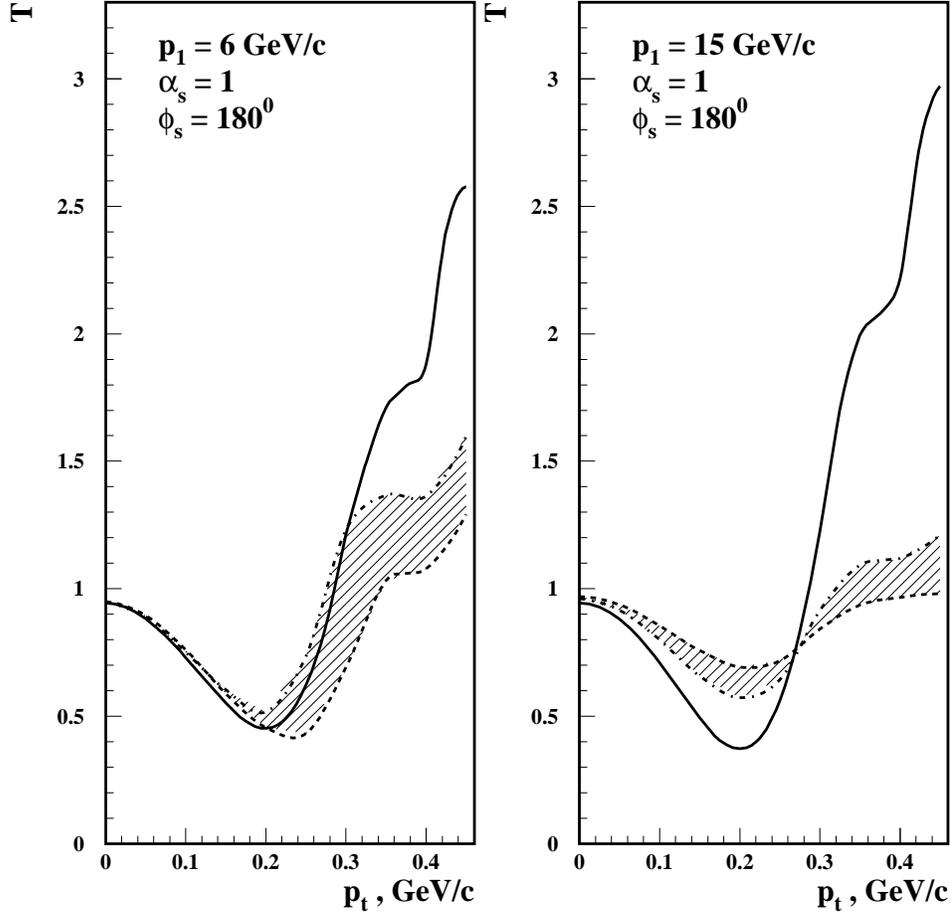,width=14cm,height=14cm}
}
\caption{ {\em The $p_t$ dependence of $T$ at $\alpha_s=1$.
The solid line is for the  elastic eikonal approximation 
which neglects color transparency 
effects. The shaded area corresponds to $T$ 
calculated within  the quantum diffusion model of CT.
Dashed and dash-dotted curves correspond to QDM calculation 
with $\Delta M^2=0.7$ and   $\Delta M^2=1.1~GeV^2$ respectively.
}}
\end{figure}

\begin{figure}[p]
\centerline{
\epsfig{file=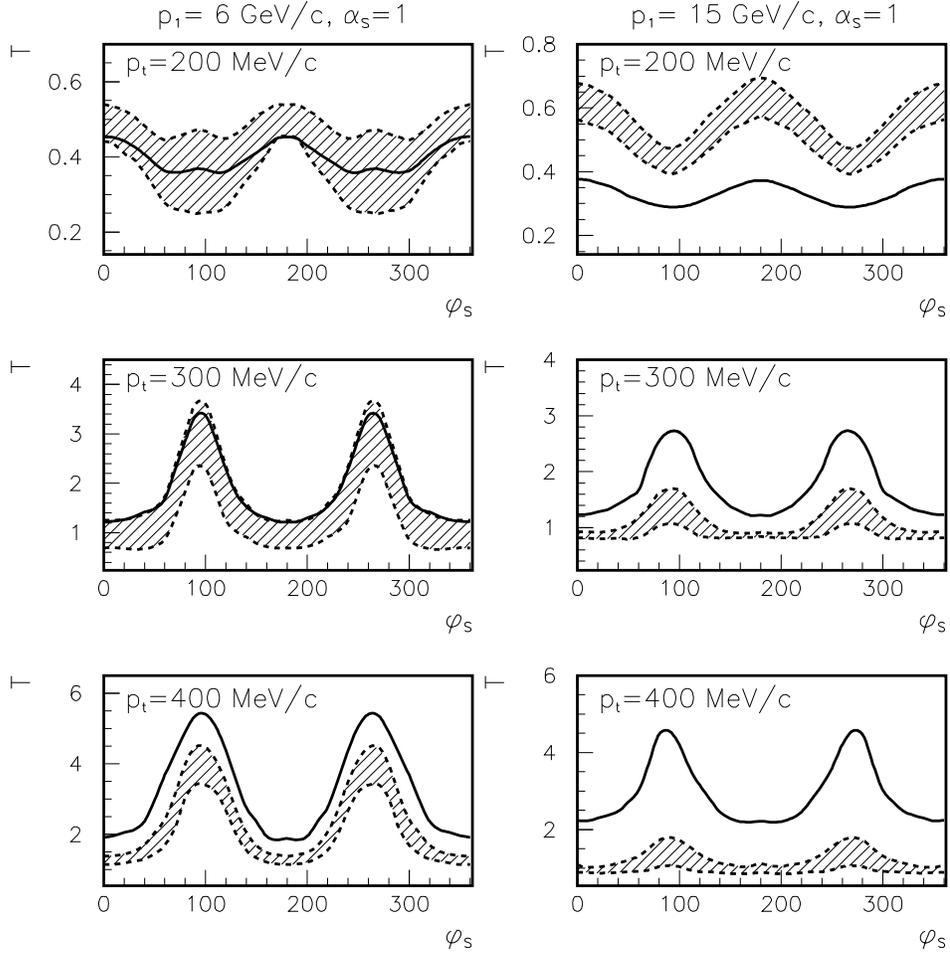,width=14cm,height=14cm}
}
\caption{ {\em  The $\phi_s$ dependence of  $T$ at  
$p_t=0.2~GeV/c$, $p_t=0.3~GeV/c$ and $p_t=0.4~GeV/c$ and 
$\alpha_s=1$. 
The solid line is the elastic eikonal approximation. 
The shaded area presents $T$ 
including CT within QDM. The parameter
$\Delta M^2$ lies in the  range $0.7-1.1~GeV^2$.}}
\end{figure}

\begin{figure}[p]
\centerline{
\epsfig{file=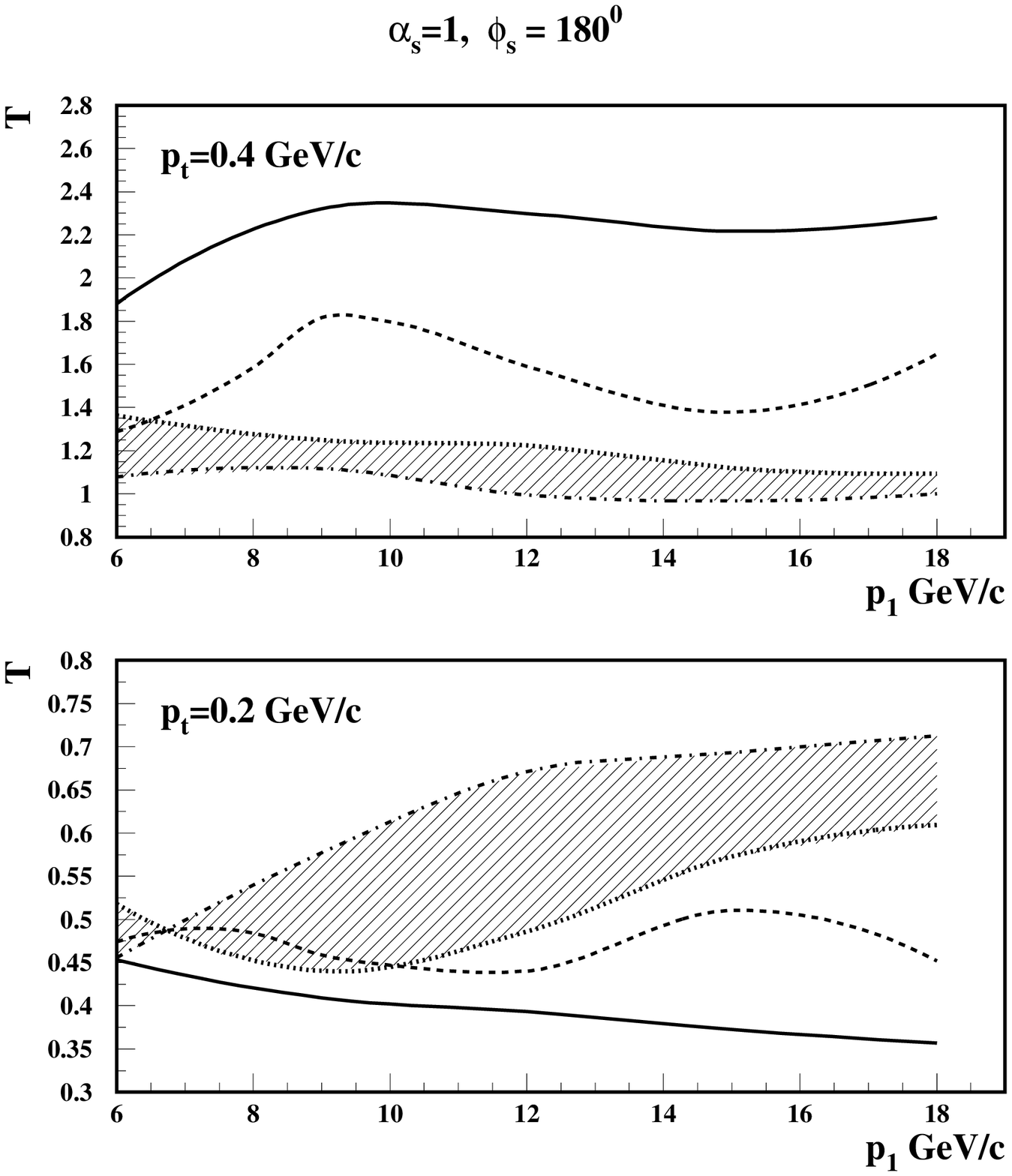,width=14cm,height=14cm}
}
\caption{ {\em $T$ as a function of 
projectile momentum $p_1$, at  
$p_t=0.2~GeV/c$ and $p_t=0.4~GeV/c$.
The solid line is the elastic eikonal approximation. 
The shaded area presents $T$ 
including QDM prediction of CT 
within quantum diffusion model with  the parameter
$\Delta M^2$  within the range $0.7-1.1~GeV^2$.
Dash-doted line - QDM prediction with $\Delta M^2=0.7~GeV^2$, dotted
line - $\Delta M^2=1.1~GeV^2$. Dashed line corresponds to prediction CT
within the model which accounts for interference between large and 
small size configurations in $pp$ hard scattering amplitude.}}
\end{figure}

\begin{figure}[p]
\centerline{
\epsfig{file=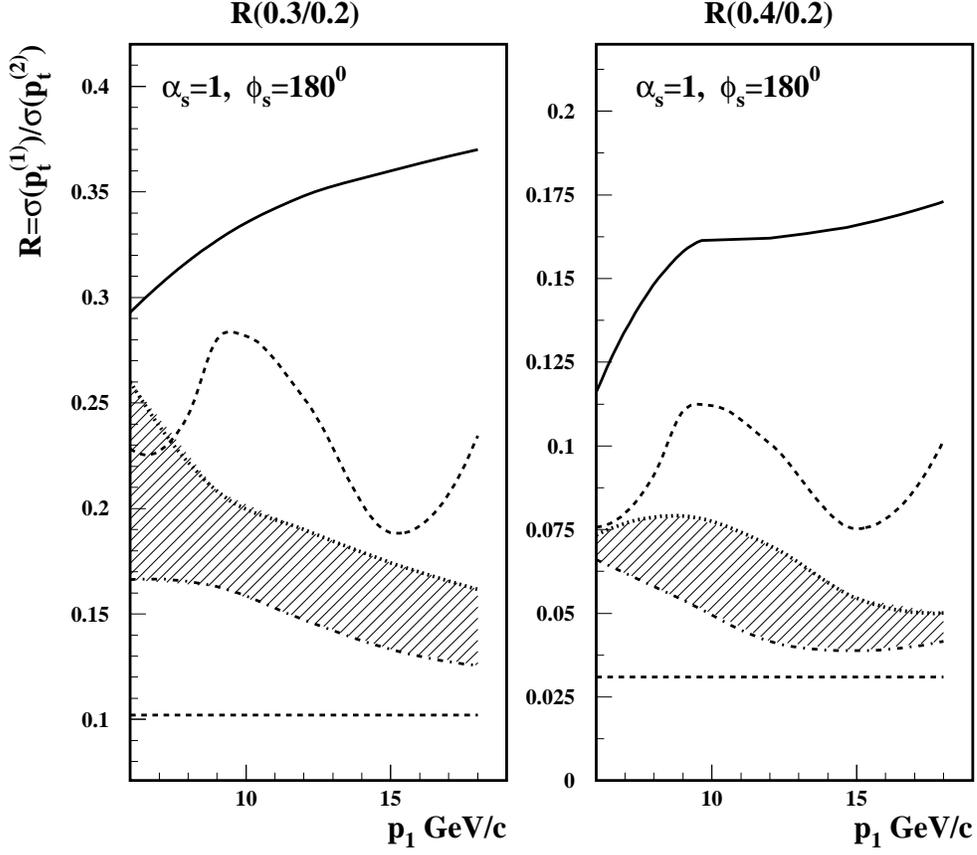,width=14cm,height=14cm}}
\caption{ {\em  The $p_1$ dependence of the ratio 
$R={\sigma(p^{(1)}_{st}\over \sigma(p^{(2)}_{st}}$ at $\alpha_s=1$.
The solid line is the elastic eikonal approximation. 
The shaded area is the the $R$  
calculated within quantum diffusion model of CT with 
$\Delta M^2$ parameter in the range $0.7-1.1~GeV^2$. The definitions
of curves are same as in Fig.8}}
\end{figure}

\begin{figure}[p]
\centerline{
\epsfig{file=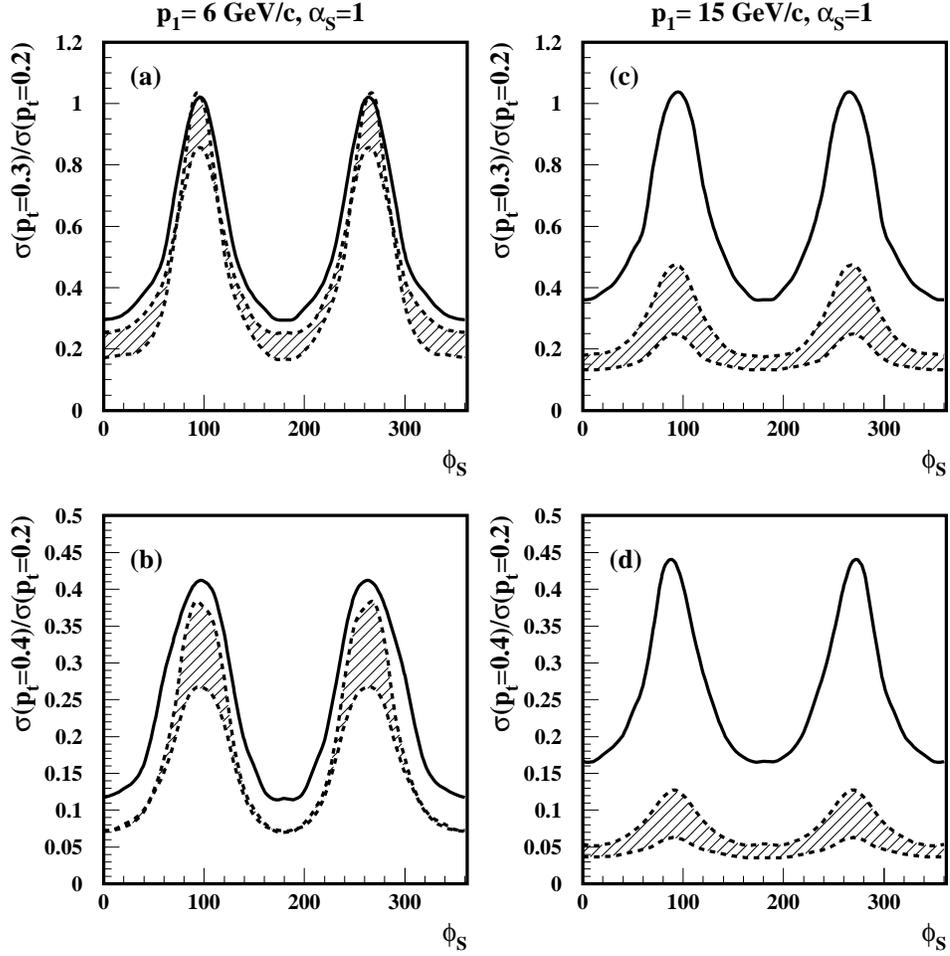,width=14cm,height=14cm}
}
\caption{ {\em  The ratio $R={\sigma(\phi_s={\pi \over 2})\over
\sigma(\phi_s=\pi)}$ at $\alpha_s=1$ as a function of the spectator
azimuthal angle $\phi_s$ 
The solid line is the elastic eikonal approximation. 
The shaded area includes CT, calculated in the QDM,  with 
$\Delta M^2$ parameter in the range $0.7-1.1~GeV^2$.}}.
\end{figure}

 \begin{figure}[p]
\centerline{
\epsfig{file=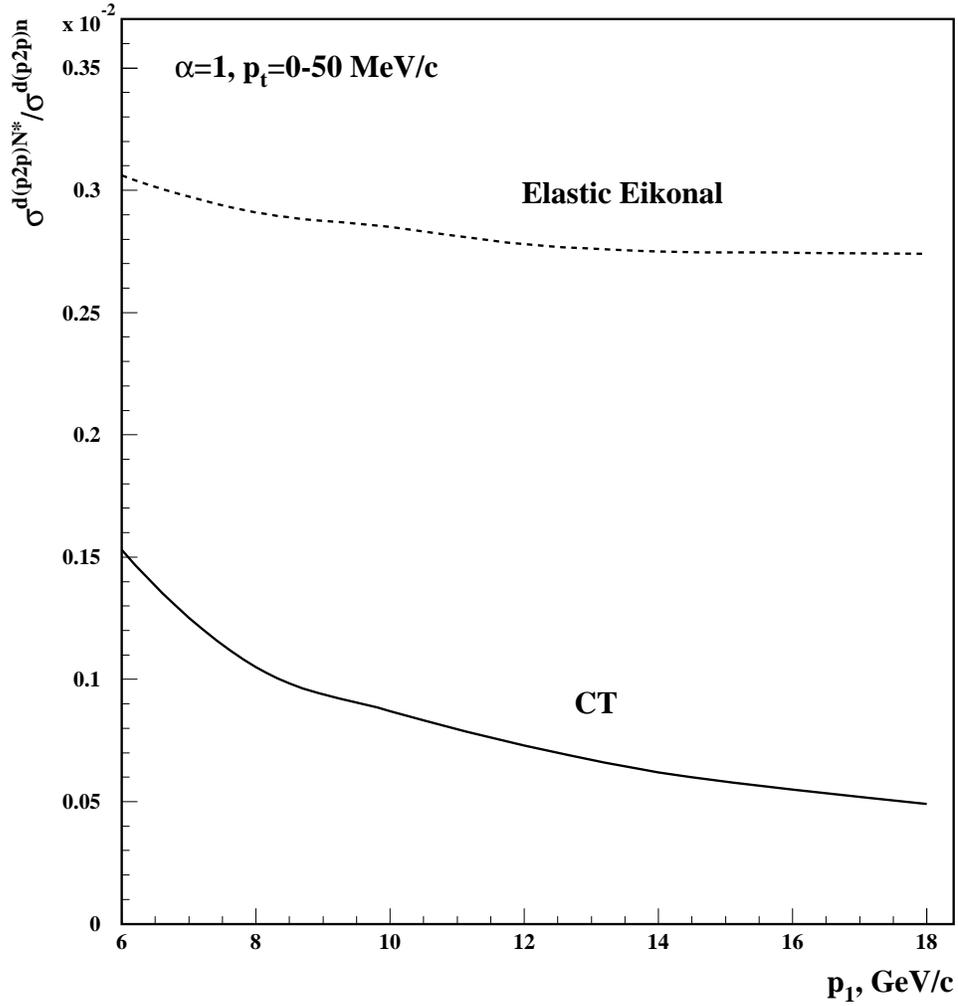,width=14cm,height=14cm}
}
\caption{ {\em The energy dependence of the ratio of the cross
sections for the reactions $d(p,2p)n$ and $d(p,2p)N^*$.
 The calculation was done for the elastic eikonal 
approximation (dashed line) and within the QDM model of CT (solid line), 
with $\Delta M^2=0.7$ GeV$^2$.}}
\end{figure}

\begin{figure}[p]
\centerline{
\epsfig{file=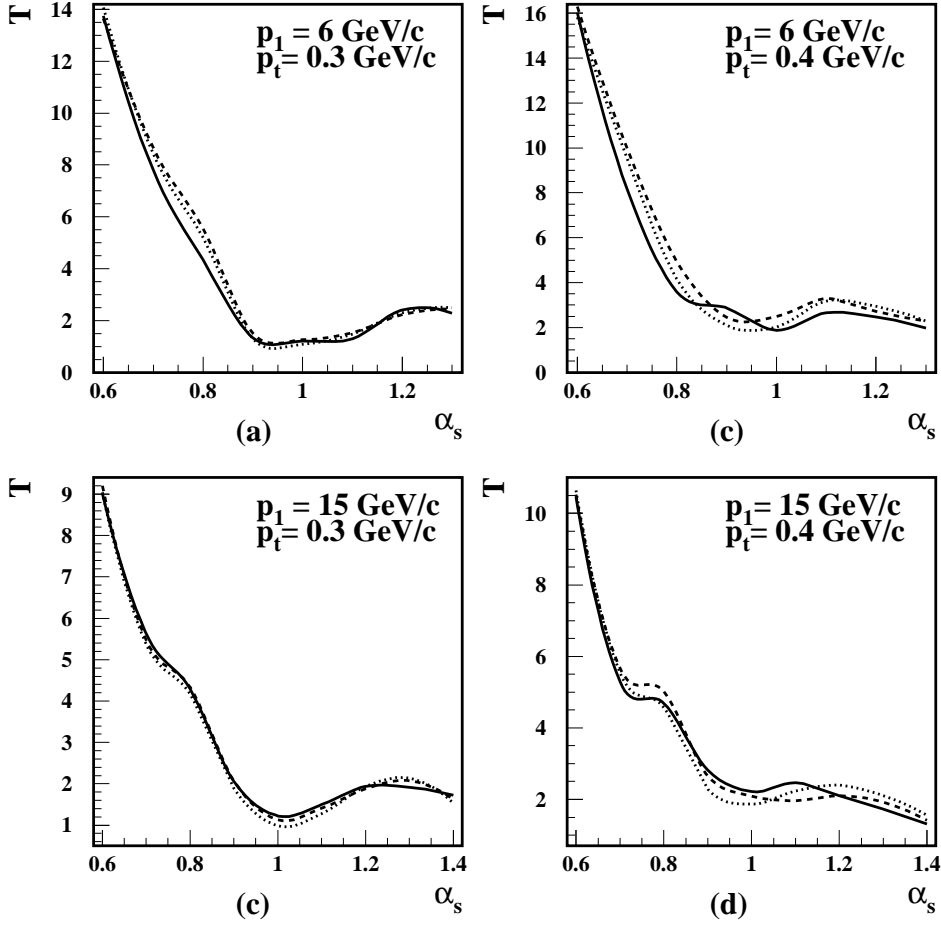,width=14cm,height=14cm}
}
\caption{ {\em $T$ as a function of $\alpha_s$   for  different $p_t$.
The solid lines is for  the complete factorization, the dashed and 
dash-dotted lines are the two calculations which do not
have the factorization as 
explained in Appendix B. }} 
\end{figure}

\end{document}